\documentclass[a4paper, 10pt]{article}
\pdfoutput=1

\usepackage{amsmath, amssymb, bm}
\usepackage{mathtools}  
\usepackage[top=1in, bottom=1.2in, left=0.5in, right=0.5in]{geometry}
\usepackage[]{graphicx}
\usepackage[font=scriptsize,labelfont=bf]{caption}
\usepackage[caption=false]{subfig}
\usepackage{textcomp}
\usepackage{authblk}
\usepackage{hyperref}

\newcommand{\bs}{\boldsymbol}

\begin{document}

\title{\Large{\bf{On the photon-pseudoscalar particle mixing in media and external fields}}}
\author{Damian Ejlli}

\affil{\emph{\normalsize{School of Physics and Astronomy, Cardiff University, The Parade, Cardiff CF24 3AA, United Kingdom}}}

\date{}

\maketitle

\begin{abstract}
In this work, I study the mixing of photons with pseudoscalar particles and vice-versa in the presence of an external magnetic field and a pseudoscalar field. I solve exactly for the first time in the literature the equations of motion of the electromagnetic field coupled with a pseudoscalar field in the presence of a constant magnetic field with arbitrary direction with respect to the direction of propagation of the fields in vacuum. In addition, I also solve exactly the equations of motion in a magnetized plasma/gas for perpendicular propagation with respect to the external magnetic field. By finding exact solutions to the equations of motion, I find exact expressions for the transition efficiencies of photons into pseudoscalar particles in different situations. The expressions of the transition efficiencies generalize and correct those previously found in the literature by using approximate WKB methods on solving the equations of motion. In the case when the direction of propagation of fields with respect to the external magnetic field is not perpendicular, a longitudinal state of the electromagnetic field is generated even in a magnetized vacuum. The appearance of the longitudinal electric field state could be used for laboratory searches of pseudoscalar particles such as the axion and/or axion-like particles.
\end{abstract}

\section{Introduction}
\label{sec:1}

One of the most interesting effects that appear in electromagnetism is the interaction of the electromagnetic field with external prescribed fields in media or a vacuum. Depending on the type of external field and on the type of medium, several effects such as birefringence and dichroism may manifest. Usually, if there is present only an external electric field, the effects that manifest are called electro-optic effects and in the case when it is present only an external magnetic field, the effects that manifest are called magneto-optic effects. Among the magneto-optic effects, one of the most important is the mixing of the electromagnetic field with a pseudoscalar field in an external magnetic field. This type of interaction subjects the electromagnetic field to birefringence and/or dichroism effects that are in principle observable depending on the values of the parameters that enter the interaction Lagrangian between the electromagnetic field, the background magnetic field, and the pseudoscalar field.

In the last four decades, there have been several efforts to study and possibly detect, the axion or axion-like particles, that are theorized pseudoscalar particles that in principle can solve the strong CP problem in QCD\cite{Kim:1979if}.
These pseudoscalar particles are expected to be very light and to interact very weakly with matter and fields and so far their detection has been quite elusive in laboratory experiments. One of the most promising ways to detect these pseudoscalar particles is to use their coupling with the electromagnetic field that usually is represented with a vertex Feynman diagram where two photons lines connect with a pseudoscalar field line. The first theoretical studies that proposed to use the two-photon coupling with the pseudoscalar field, suggest using the interaction of an electromagnetic wave with an external prescribed magnetic field that eventually leads to a partial or complete transformation of the electromagnetic wave into a propagating pseudoscalar field in the same direction of the propagating electromagnetic wave \cite{Sikivie:1983ip}-\cite{Raffelt88}. The effect of the partial or complete transformation of the electromagnetic wave into a propagating  pseudoscalar field makes it possible that the incident electromagnetic waves manifest birefringence and dichroism effects, which magnitude, depends on the incident electromagnetic wave frequency, external magnetic field strength, and direction, and on the type of medium where the interacting fields propagate in.

Based on the two-photon coupling of the pseudoscalar field, in the recent years there have been several experimental efforts to detect axions and axion-like particles in a laboratory when usually laser beams interact with a transverse magnetic field and after the effects of this interaction on the incident electromagnetic wave is studied \cite{Cameron:1993mr}. Among these experiments the so-called ``light shining through a wall" experiment aims to convert the incident electromagnetic wave into weakly interacting particles (axions, axion-like particles, etc), in a constant transverse magnetic field, before a wall fixed at a given distance, and after when these weakly interacting particles exit the wall, try to re-convert these weakly interacting particles into photons in a constant and transverse magnetic field. Another possibility to use the two-photon coupling to detect axions or axion-like particles is to measure any possible rotation of the incident electromagnetic wave polarization plane when it exits the external magnetic field \cite{Zavattini:2005tm}. So far, all these different types of experiments have not detected any axion or axion-like particle or any other weakly interacting particle but only upper or lower limits on the mass and/or coupling constant to photons of these particles have been set.

So far, on the theoretical side of the two-photon coupling of the pseudoscalar field, all theoretical calculations have been based on approximate methods to solve the coupled equations of motion of the electromagnetic field with the pseudoscalar field. Typically, the coupled field interaction is reduced to a set of coupled differential equations and they are solved by using a WKB method to linearised the equations of motion and after use a semi-classical approach to calculate transition probabilities and the rotation angle of the polarization plane or the generated ellipticity of the incident electromagnetic wave. Also, in these theoretical studies \cite{Sikivie:1983ip}-\cite{Raffelt88}, which so far have been the main ones in this field, the magnetic field is assumed to be constant and it can be either perpendicular or it can have a longitudinal component with respect to the direction of propagation of the fields. For example, in Ref. \cite{Maiani:1986md}, the external magnetic field is assumed to be constant and it has a longitudinal component with respect to the direction of propagation of fields and possible matter effects on the electromagnetic wave are completely neglected. Then the coupled photon-pseudoscalar particle equations of motion are solved in momentum space. In Ref. \cite{Raffelt88}, the coupled photon-pseudoscalar particle equations of motions are linearised and after are solved for only a transverse external magnetic field in presence of matter for only relativistic photons and pseudoscalar particles. Some perturbative solutions and possible matter effects in the context of dark energy and light shining through wall experiment have also been studied in Ref. \cite{Das:2004qka} -\cite{Adler:2008gk}.

One important fact about the photon-pseudoscalar particle equations of motion is that they have never been solved exactly (I will discuss in next sections what is meant with ``exact solution'') in vacuum/matter and/or for non-perpendicular propagation of the fields with respect to the external magnetic field. An important issue that is usually ignored or neglected in these theoretical studies is that in the case when the magnetic field has a longitudinal component with respect to the direction of propagation of the fields, a longitudinal electric state is excited either the mixing happens in a magnetized vacuum or magnetized matter such as plasmas or gases. Some studies have explored the appearance of the longitudinal electromagnetic state in plasma\cite{Mikheev:1998bg} and in the context of dark matter \cite{Millar:2017eoc}-  \cite{Caputo:2020quz} where the axion-photon equation of motion have been solved approximately in momentum space for an isotropic and homogeneous medium and no explicit calculation of the field evolutions have been presented. The appearance of the longitudinal electric state could play an important role in helping the detection of axions or other similar particles. The main goal of this work is to go beyond the initial theoretical studies \cite{Sikivie:1983ip}-\cite{Raffelt88}, and solve exactly the coupled equations of motion of the photon-pseudoscalar particle (such as axions and axion-like particles) mixing in an external constant magnetic field by using the same approach as presented in Ref. \cite{Ejlli:2020fpt} and find useful quantities such as transition efficiencies for arbitrary particle energies in magnetized vacuum or media. We will see that in some cases when the transition efficiency of photons into pseudoscalar particles is equal to zero by using the approximate methods discussed above, in the case when the transition probability is found by exactly solving the equations of motion, it contains other additional terms that can be relevant depending on the situation. 

This work is organized as follows: In Sec. \ref{sec:2}, I formulate the problem of the photon-pseudoscalar particle mixing in a constant magnetic  and pseudoscalar field. In Sec. \ref{sec:3}, I exactly solve the equations of motion in a magnetized vacuum only for an arbitrary direction of the external magnetic field. In Sec. \ref{sec:4}, I exactly solve the equations of motion in a magnetized plasma or gas for a transverse magnetic field only. In Sec. \ref{sec:5}, I find an exact expression for the transition probability of photons into pseudoscalar particles and find the strength of the electric field of the excited longitudinal photon state. In Sec. \ref{sec:6}, I conclude. In this work I use the metric with signature $\eta_{\mu\nu}=\text{diag}[1, -1, -1, -1]$ and work with the rationalized Lorentz-Heaviside natural units ($k_B=\hbar=c=\varepsilon_0=\mu_0=1$) with $e^2=4\pi \alpha$.

\section{Mixing in an external magnetic and pseudoscalar fields}
\label{sec:2}

In this section we derive the equations of motion for the photon and pseudoscalar fields propagating in a magnetized  medium. To start with, we write the effective action of the electromagnetic and pseudoscalar fields in flat spacetime as
\begin{equation}\label{tot-act}
\mathcal S_\text{eff} =\int d^4x \left(-\frac{1}{4} F_{\mu\nu}F^{\mu\nu}-\frac{1}{2}\int d^4x^\prime\,A_\mu (x)\Pi^{\mu\nu}(x, x^{\prime})A_\nu(x^{\prime})+\frac{1}{2}\partial_\mu\phi \partial^\mu \phi -\frac{1}{2}m_\phi^2\,\phi^2+\frac{g_{\phi\gamma}}{4}\phi\,F_{\mu\nu}\tilde F^{\mu\nu}\right),
\end{equation}
where $F_{\mu\nu}$ is the total electromagnetic field tensor, $\Pi^{\mu\nu}$ is the photon polarization tensor in a medium, $\phi$ is the pseudoscalar field, $m_\phi$ is the mass of the pseudoscalar field, $g_{\phi\gamma}$ is the coupling constant of photons with pseudoscalar particles, and $A^\mu$ is the photon vector potential. By varying the action \eqref{tot-act} with respect to the electromagnetic field $A^\nu$ and pseudoscalar field $\phi$, we find the following equations of motion 
\begin{eqnarray}\label{eq-mo}
\Box A^\nu-\partial_\mu(\partial^\nu A^\mu)-\int d^4 x^\prime\, \Pi^{\nu\mu}(x, x^\prime)\,A_\mu(x^\prime) &=& g_{\phi\gamma} (\partial_\mu \phi) \tilde F^{\mu\nu},\nonumber\\
(\Box+m_\phi^2) \phi=\frac{g_{\phi\gamma}}{4}F_{\mu\nu}\tilde F^{\mu\nu},
\end{eqnarray}
where we used the other set of Maxwell's equations $\partial_\mu \tilde F^{\mu\nu}=0$ in the first equation in \eqref{eq-mo}  and $x^\mu=(t, \bs x^i)$, $x_\mu=(t, \bs x_i)$. In this work we will use the standard notations for the spatial components of a generic three dimensional vector $\bs a=(a^1, a^2, a^3)\equiv (a_x, a_y, a_z)$.

In general, the electromagnetic field tensor $F_{\mu\nu}$ is given by the sum of the field tensor of the incident photon field $f_{\mu\nu}$ and of the field tensor corresponding to the external magnetic field $\bar F_{\mu\nu}$. In most cases, the electromagnetic field tensor corresponding to the external magnetic field is the dominant term, so, $F^{\mu\nu}\simeq \bar F^{\mu\nu}$. Considering the photon propagation in an external magnetic field only, the equations of motion \eqref{eq-mo} for the components of vector potential $\bs A^i (x)$ and pseudoscalar field $\phi(x)$ in the temporal gauge\footnote{For an exhaustive discussion on the gauge conditions in electromagnetism, including the temporal gauge, and their interconnection see Ref. \cite{Jackson:2002rj}.} $A^0(x)=0$ are
\begin{eqnarray}\label{eq-A-ph}
\partial_i\partial^t \bs A^i (x)  & = & - g_{\phi\gamma} \partial_i\phi(x) \bar{\bs B}^i - \int d^4 x^\prime\,\Pi^{0i}(x, x^\prime) \bs A_i(x^\prime),\nonumber\\
(\partial_t^2-\nabla^2) \bs A^i(x) + \partial_i\partial_j\bs A^j(x) - \int d^4 x^\prime\,\Pi^{ij}(x, x^\prime) \bs A_j(x^\prime) &=&-g_{\phi\gamma} \partial_t\phi(x) \bar{\bs B}^i,\nonumber\\
(\partial_t^2-\nabla^2 + m_\phi^2)\phi(x) &=& g_{\phi\gamma}\partial_t\bs A_i(x)\cdot\bar{\bs B}^i,
\end{eqnarray}
where $\bar{\bs B}^i$ are the components of the external magnetic field vector and we used the fact that only the dominant term in $F_{\mu\nu}\tilde F^{\mu\nu} \simeq -4 \bs E_\gamma\cdot \bar{\bs B}$ is kept with $\tilde F^{0i} \simeq -\bar{\bs B}^i$ and $\bar{\bs E}=0$ and $\partial^i=-\partial_i$. Let us now expand the propagating electromagnetic field $\bs A^i$ and pseudo-scalar field $\phi$ as Fourier integrals in $\omega$
\begin{align}\label{field-expansion}
\bs A^i({\bs x}, t) &=\sum_{\lambda=x, y, z} \int_{-\infty}^{+\infty}\frac{d\omega}{2\pi} e_\lambda^{i}(\hat{\bs n}) A_{\lambda}({\bs x, \omega})e^{-i \omega t},\quad  \phi(\bs x, t)= \int_{-\infty}^{+\infty}\frac{d\omega}{2\pi}   \phi(\bs x, \omega)e^{-i \omega t},\end{align}
where $e_\lambda^i$ is the electromagnetic wave polarization vector that depends on the direction of propagation $\hat{\bs n}$.  In addition, we have that $A_{\lambda}({\bs x, \omega})= A_{\lambda}^*({\bs x, -\omega})$ and $\phi(\bs x, \omega)= \phi^*(\bs x, -\omega)$ in order to ensure that $\bs A^i({\bs x}, t) $ and $\phi(\bs x, t)$ are real functions. We may note from the system \eqref{eq-A-ph} that the presence of a magnetized medium makes it possible the appearance of the longitudinal electromagnetic field component.  

In what follows, we assume that the medium where electromagnetic waves propagate to be homogeneous and anisotropic due to the presence of the magnetic field. In this case the response of the medium is linear where $\Pi^{ij}(x, x^\prime)=\Pi^{ij}(x-x^\prime)$. We also assume that the effects of the medium on the propagation of electromagnetic waves to be local in space, namely $\Pi^{\mu\nu}(\bs x-\bs x^\prime, t-t^\prime) \simeq \Pi^{\mu\nu}(t-t^\prime)\delta^3(\bs x-\bs x^\prime)$. This approximation is quite accurate in most simple media such as gases and plasma, see Ref. \cite{Landau:1987gn} for details. The case when the medium is spatially dispersive is not studied in this work since for a cold collisionless plasma, as we study in Sec. \ref{sec:4}, the spatial dispersion can be safely neglected \cite{Shafranov:1958}.  In general, the photon polarization tensor is a complex valued quantity, where usually its hermitian part includes dispersive phenomena in media, while its anti-hermitian  part includes absorption/dissipative phenomena in media. In what follows we assume $\Pi^{ij}$ to be a complex valued hermitian quantity $\Pi_{ij}=\Pi_{ji}^*$ which includes dispersive phenomena only in a magnetized medium. The effects of the pseudoscalar field on the electromagnetic field are not included in $\Pi^{ij}$ but appear as a source term in the Maxwell equations in \eqref{eq-A-ph}. Dissipative/absorption  processes of the electromagnetic radiation that may be included in $\Pi^{ij}$ in addition to the pseudoscalar source term are not studied in this work.

Without any loss of generality, consider the case when the electromagnetic and the pseudoscalar fields propagate in a given coordinate system along the $\hat{\bs z}$ axis, namely $\hat{\bs n}=\hat{\bs z}$. We assume that all propagating fields depend on the $z$ coordinate only. By using the field expansions in \eqref{field-expansion} in the system \eqref{eq-A-ph} and making use of the assumptions about $\Pi^{ij}$, the system \eqref{eq-A-ph} becomes 
\begin{align}\label{eq-A-ph-1}
\partial_i\bs A^i (z, \omega)  & =  - \frac{ i g_{\phi\gamma}}{\omega} \partial_i\phi(z, \omega) \bar{\bs B}^i  - \frac{i \Pi^{0i}(\omega)}{\omega} \bs A_i(z, \omega) \nonumber\\
(\omega^2 + \partial_z^2) \bs A^i(z, \omega) - \partial_i\partial_j\bs A^j(z, \omega) + \Pi^{ij}(\omega) \bs A_j(z, \omega) &=-i\omega g_{\phi\gamma} \phi(z, \omega) \bar{\bs B}^i,\nonumber\\
(\omega^2 + \partial_z^2 - m_\phi^2)\phi(z, \omega) &= i\omega g_{\phi\gamma}\bs A_i(z, \omega)\cdot\bar{\bs B}^i.
\end{align}
In terms of the field components $\bs A^i=(A^1, A^2, A^3)\equiv (A_x, A_y, A_z)$, with $\bs A_i=-\bs A^i$, we get the following system of partial differential equations 
\begin{gather}
\partial_z A_z(z, \omega) = - \frac{ig_{\phi\gamma}}{\omega}\partial_z\phi(z, \omega)\bar B_z + \frac{i\Pi^{0x}(\omega)}{\omega} A_x(z, \omega)  + \frac{i\Pi^{0y}(\omega)}{\omega} A_y(z, \omega) + \frac{i\Pi^{0z}(\omega)}{\omega} A_z(z, \omega), \nonumber\\
(\omega^2 + \partial_z^2) A_x(z, \omega)  - \Pi^{xx}(\omega) A_x(z, \omega) - \Pi^{xy}(\omega) A_y(z, \omega)- \Pi^{xz}(\omega) A_z(z, \omega ) = -i\omega g_{\phi\gamma} \phi(z, \omega) \bar{ B}_x, \nonumber\\
(\omega^2 + \partial_z^2) A_y(z, \omega)  - \Pi^{yx}(\omega) A_x(z, \omega) - \Pi^{yy}(\omega) A_y(z, \omega)- \Pi^{yz}(\omega) A_z(z, \omega) =-i\omega g_{\phi\gamma} \phi(z, \omega) \bar{ B}_y, \nonumber\\
\left[ \omega^2 - \Pi^{zz}(\omega) \right] A_z (z, \omega) - \Pi^{zx}(\omega) A_x(z, \omega) - \Pi^{zy}(\omega) A_y(z, \omega) + i\omega g_{\phi\gamma} \phi(z, \omega) \bar{ B}_z=0, \nonumber\\
(\omega^2 + \partial_z^2 - m_\phi^2)\phi(z, \omega)  = i\omega g_{\phi\gamma} \left[ A_x(z, \omega) \bar{B}_x + A_y(z, \omega) \bar{B}_y + A_z(z, \omega) \bar{B}_z \right]. \label{eq-A-ph-2}
\end{gather}

The system of partial differential equations in \eqref{eq-A-ph-2} is in its final form\footnote{The temporal elements of the photon polarization tensor in \eqref{eq-A-ph-2} can be expressed in terms of the spatial components $\Pi_{ij}$ if we go in momentum space. Indeed, it can be shown that for a linear medium, the elements of the photon polarization tensor in momentum space must obey to the charge-continuity equation $k_\mu J^\mu(\bs k, \omega)=0$ and the gauge invariance transformation of the fields. In this case the photon polarization tensor must satisfy the following conditions in momentum space $k_\mu \Pi^{\mu\nu}(\bs k, \omega)=0$ and $k_\nu \Pi^{\mu\nu}(\bs k, \omega)=0$.} and it describes the mixing of the electromagnetic field with the pseudoscalar field in a magnetized medium. We may note that the fourth equation in \eqref{eq-A-ph-2}  is in reality a constraint condition that all fields must satisfy for any $z, \omega$ and it is not a dynamical equation to be solved with the remaining set of equations. To solve the system \eqref{eq-A-ph-2}, we need first to entirely reduce it to a system of the first order of linear differential equations. To achieve it, let us make the following definitions:
\begin{gather}
A_x(z, \omega)  \equiv x_1(z, \omega), \quad \partial_z A_x(z, \omega)  \equiv x_2(z, \omega), \quad A_y(z, \omega)  \equiv x_3(z, \omega), \quad \partial_z A_y(z, \omega)  \equiv x_4(z, \omega), \nonumber \\A_z(z, \omega)  \equiv x_5(z, \omega), \quad \phi(z, \omega)  \equiv x_6(z, \omega), \quad \partial_z \phi(z, \omega)  \equiv x_7(z, \omega).
\end{gather}

Now we define the column field $X(z, \omega) = \left[ x_1(z, \omega) , x_2(z, \omega) , x_3(z, \omega) , x_4(z, \omega) , x_5(z, \omega) , x_6(z, \omega) , x_7(z, \omega)  \right]^{\text{T}}$ where the symbol ($\text{T}$) means the transpose of a given field with components. In this case, the system of differential equations \eqref{eq-A-ph-2} can be written as
\begin{equation}\label{system-1}
\partial_z X(z, \omega) = M (\omega, \bar B_x, \bar B_y, \bar B_z)X(z, \omega),
\end{equation}
where the matrix $M$ is given by
\begin{equation}\label{matrix-M}
M=\begin{pmatrix}
0 & 1 & 0 & 0 & 0 & 0 & 0 \\
\Pi^{xx}(\omega) -\omega^2 & 0 & \Pi^{xy}(\omega) & 0 & \Pi^{xz}(\omega) & -\sigma \bar B_x & 0 \\
0 & 0 & 0 & 1 & 0 & 0 & 0 \\
\Pi^{yx}(\omega) & 0 &\Pi^{yy}(\omega) -\omega^2 & 0 & \Pi^{yz}(\omega) & -\sigma \bar B_y & 0\\
\frac{i\Pi^{0x}(\omega)}{\omega} & 0  &  \frac{i\Pi^{0y}(\omega)}{\omega}  & 0 &  \frac{i\Pi^{0z}(\omega)}{\omega} & 0 & -\sigma\bar B_z/\omega^2\\
0 & 0 & 0 & 0 & 0 & 0 & 1 \\
\sigma \bar B_x &  0 & \sigma \bar B_y & 0 & \sigma \bar B_z & m_\phi^2 -\omega^2 & 0 \\
   \end{pmatrix},      
\end{equation}
where we have defined the complex quantity $\sigma \equiv i\omega g_{\phi\gamma}$.
Since the matrix $M$ does not explicitly depend on $z$, it commutes with itself at different positions, namely $[M, M]=0$ for every $z$. In this case, the general solution of \eqref{system-1} is given by
\begin{equation}\label{sol-eq-mot}
X(z, \omega) = \exp{\left[\int_{z_i}^z dz^\prime M(\omega,  \bar B_x, \bar B_y, \bar B_z)\right]} X(z_i, \omega),
\end{equation}
where $z_i$ is the initial position where the interaction of the propagating fields with the external magnetic field takes place. As far as the matrix exponential is concerned, the solutions of the equations of motion is simply reduced on calculating the exponential of $(z-z_i) M$.

In what follows we concentrate on finding a solution to the equations of motion \eqref{system-1}. In this work with ``exact solution" of the equations of motion, we mean the exact solution of \eqref{system-1} in some particular cases such as in magnetized vacuum and in media. The equations of motion \eqref{system-1} have never been solved exactly before in the situations that we study in this work but only by using an approximate WKB method to linearize the equations \eqref{system-1} and after solve them in some particular cases. Evidently one has to keep in mind some of the approximations used to derive \eqref{system-1}, namely a spatially local polarization tensor $\Pi^{\mu\nu}$ and only the linear terms of the (weak with respect to $\bar{\bs B}$) incident electromagnetic field are kept in \eqref{eq-A-ph}. Therefore, with ``exact solution" of the equations of motion we essentially mean the exact solution of the equations in \eqref{system-1} only. Obviously, one has to distinguish between the exact solution of the nonlinear equations in \eqref{eq-mo} for arbitrary media, which is beyond the scope of this work, and the solution of the system of equations \eqref{eq-A-ph-2} or \eqref{system-1}.

\section{Solution of equations of motion in an external magnetic field only}
\label{sec:3}

Consider now the case of electromagnetic waves (photons) mixing with a pseudo-scalar field in the presence of an external magnetic field only. With this statement, we mean that there is not matter (gas and/or plasma) present in the laboratory but only a magnetic field, so, we can treat the mixing problem as if it happens in a magnetized vacuum. In this case, we can approximate all the elements of the photon polarization tensor to zero, $\Pi^{\mu\nu}(\omega)\simeq 0$. Of course, this approximation is quite accurate in the case when the strength of the external magnetic field is reasonably weak in comparison with the strength of the critical magnetic field $B_c$. This essentially means that for laboratory magnetic field strengths we can neglect as well the effect of vacuum polarization in the presence of the external magnetic field. Under this hypothesis, we can neglect all the elements of the photon polarization tensor in the matrix $M$. In this case, the matrix $M$ is quite simplified and we can calculate the matrix exponential $(z-z_i) M$ analytically if $M$ is a diagonalizable matrix\footnote{In this work we calculate the matrix exponential of $(z-z_i) M$ only in the case when the matrix $M$ is diagonalizable. In this case we eventually have to constraint the parameter space only for those values of the parameters in which $M$ is diagonalizable, as we will see in what follows. The case when $M$ is not diagonalizable, the matrix exponential of $(z-z_i) M$ can be still calculated analytically with other methods that are beyond the scope of this work.}. The matrix $M$ has seven distinct eigenvalues and it is diagonalizable, namely $M=P D P^{-1}$ where $P$ is the matrix formed with the eigenvectors of $M$ and $D$ is a diagonal matrix formed with the eigenvalues of $M$. Since the matrix $M$ is diagonalizable, we have that $\exp[(z-z_i)M]= P\exp[(z-z_i)D] P^{-1}$. The eigenvalues $\lambda_i$ and eigenvectors $v_i$ of $M$ for $i=1,2,3,4,5,6, 7$ are 
\begin{equation*}
\begin{gathered}
\lambda_1= 0,\, \lambda_2=-i\omega,\, \lambda_3=i\omega,\, \lambda_4=-\mathcal C_-/(\sqrt{2}\omega^2),\, \lambda_5=
\mathcal C_-/(\sqrt{2}\omega^2),\, \lambda_6= -\mathcal C_+/(\sqrt{2}\omega^2),\, \lambda_7= \mathcal C_+/(\sqrt{2}\omega^2),\\
v_1= \left[-\sigma \bar B_x/\omega^2, 0, -\sigma \bar B_y/\omega^2, 0, \frac{\omega^4 + \sigma^2 B_T^2- \omega^2 m_\phi^2}{\sigma \omega^2 \bar B_z}, 1, 0\right]^\text{T}, \, v_2=\left[-i\bar B_y/(\omega \bar B_x), -\bar B_y/\bar B_x, i/\omega, 1, 0, 0, 0\right]^\text{T},\\
v_3=\left[i\bar B_y/(\omega \bar B_x), -\bar B_y/\bar B_x, -i/\omega, 1, 0, 0, 0 \right]^\text{T},\, v_4= \left[  - \frac{2\sqrt{2}\sigma\omega^6\bar B_x}{\mathcal C_- \mathcal G_+},\,     \frac{2\sigma\omega^4\bar B_x}{\mathcal G_+},\, - \frac{2\sqrt{2}\sigma\omega^6\bar B_y}{\mathcal C_- \mathcal G_+},\,   \frac{2\sigma\omega^4\bar B_y}{\mathcal G_+},\,
  \frac{\sqrt{2}\sigma\bar B_z}{\mathcal C_-},\,  - \frac{\sqrt{2}\omega^2}{\mathcal C_-},\,  1 \right]^\text{T},\\
  v_5= \left[   \frac{2\sqrt{2}\sigma\omega^6\bar B_x}{\mathcal C_- \mathcal G_+},\,     \frac{2\sigma\omega^4\bar B_x}{\mathcal G_+},\,  \frac{2\sqrt{2}\sigma\omega^6\bar B_y}{\mathcal C_- \mathcal G_+},\,   \frac{2\sigma\omega^4\bar B_y}{\mathcal G_+},\,
 - \frac{\sqrt{2}\sigma\bar B_z}{\mathcal C_-},\,   \frac{\sqrt{2}\omega^2}{\mathcal C_-},\,  1 \right]^\text{T}, \\ v_6= \left[  - \frac{2\sqrt{2}\sigma\omega^6\bar B_x}{\mathcal C_+ \mathcal G_-},\,     \frac{2\sigma\omega^4\bar B_x}{\mathcal G_-},\, - \frac{2\sqrt{2}\sigma\omega^6\bar B_y}{\mathcal C_+ \mathcal G_-},\,   \frac{2\sigma\omega^4\bar B_y}{\mathcal G_-},\,
  \frac{\sqrt{2}\sigma\bar B_z}{\mathcal C_+},\,  - \frac{\sqrt{2}\omega^2}{\mathcal C_+},\,  1 \right]^\text{T},\\
   v_7= \left[   \frac{2\sqrt{2}\sigma\omega^6\bar B_x}{\mathcal C_+ \mathcal G_-},\,     \frac{2\sigma\omega^4\bar B_x}{\mathcal G_-},\,  \frac{2\sqrt{2}\sigma\omega^6\bar B_y}{\mathcal C_+ \mathcal G_-},\,   \frac{2\sigma\omega^4\bar B_y}{\mathcal G_-},\,
 - \frac{\sqrt{2}\sigma\bar B_z}{\mathcal C_+},\,   \frac{\sqrt{2}\omega^2}{\mathcal C_+},\,  1 \right]^\text{T} , 
 \end{gathered}
\end{equation*}
where we have defined $\mathcal C_{\pm}$, $\mathcal G_{\pm}$ and $\mathcal F$ as follows
\begin{equation}\label{def-0}
\mathcal C_{\pm} \equiv \left[ -2\omega^6 -\sigma^2\omega^2\bar B_z^2 +\omega^4 m_\phi^2 \pm \mathcal F \right]^{1/2}, \quad \mathcal G_{\pm} \equiv \sigma^2\omega^2\bar B_z^2 - \omega^4 m_\phi^2 \pm \mathcal F, \quad \mathcal F \equiv \left[ -4\sigma^2 \omega^8 \bar B_T^2+ \omega^4 \left(\sigma^2 \bar B_z^2 -\omega^2 m_\phi^2 \right)^2 \right]^{1/2},
\end{equation}
with $\mathcal G_- - \mathcal G_+ = -2\mathcal F$, $\bar B^2= \bar B_T^2 + \bar B_z^2$ and $\bar B_T^2= \bar B_x^2 + \bar B_y^2$. The matrix $P$ formed with the eigenvectors of $M$ is invertible when its determinant is different fro zero, namely when \begin{equation*}
\text{Det}[P]= \frac{64\,i\,\sigma\,\omega^9(2 \mathcal F)^2 \bar B_T^2(\omega^4 +\sigma^2 \bar B^2 -\omega^2 m_\phi^2  )}{\mathcal C_-\mathcal C_+ (\mathcal G_-\mathcal G_+)^2\bar B_x^2\bar B_z } \neq 0.
\end{equation*}

By performing several lengthy operations on calculating the exponential in \eqref{sol-eq-mot}, we find the following solutions for electromagnetic field components $A_{x, y, z}(z, \omega)$ and pseudoscalar field $\phi(z, \omega)$ (we do not show the solutions for the field derivatives since they are unnecessary) in a magnetized vacuum 
\begin{gather*}
A_x(z, \omega)= \frac{\left[\cosh{\left( \frac{z\, \mathcal C_+ }{\sqrt{2}\omega^2}\right)} \mathcal G_+ - \cosh{\left( \frac{z\, \mathcal C_- }{\sqrt{2}\omega^2}\right)} \mathcal G_- \right] \bar B_x^2 + 2 \cos(z\omega) \mathcal F\bar B_y^2}{2\mathcal F\bar B_T^2} A_x(0, \omega)\, + \nonumber \\  \frac{\left[\sinh{\left( \frac{z\, \mathcal C_+ }{\sqrt{2}\omega^2}\right)} \mathcal G_+ \mathcal C_- - \sinh{\left( \frac{z\, \mathcal C_- }{\sqrt{2}\omega^2}\right)} \mathcal G_- \mathcal C_+ \right] \sqrt{2}\omega^3 \bar B_x^2 + 2\mathcal F \mathcal C_-\mathcal C_+ \sin(z\omega) \bar B_y^2}{2\,\omega \,\mathcal C_-\, \mathcal C_+\,\mathcal F \bar B_T^2} \partial_z A_x(0, \omega)\, +\nonumber \\ \frac{\left( \left[\cos{\left( z\omega \right)} - \cosh{\left( \frac{z\, \mathcal C_- }{\sqrt{2}\omega^2}\right)} \right] \mathcal G_- - \left[\cos{\left( z\omega \right)} - \cosh{\left( \frac{z\, \mathcal C_+ }{\sqrt{2}\omega^2}\right)} \right] \mathcal G_+ \right) \bar B_x \bar B_y}{2\,\mathcal F \bar B_T^2} A_y(0, \omega)\, + \nonumber\\
\frac{\left( \left[\sinh{\left( \frac{z\, \mathcal C_+ }{\sqrt{2}\omega^2}\right)} \mathcal G_+ \mathcal C_- - \sinh{\left( \frac{z\, \mathcal C_- }{\sqrt{2}\omega^2}\right)} \mathcal G_- \mathcal C_+ \right] \sqrt{2}\omega^3 - 2\mathcal F \mathcal C_-\mathcal C_+ \sin(z\omega) \right) \bar B_x \bar B_y}{2\,\omega \,\mathcal C_-\, \mathcal C_+\,\mathcal F \bar B_T^2} \partial_z A_y(0, \omega) + \nonumber \\
\frac{\left( \left[1 - \cosh{\left( \frac{z\, \mathcal C_- }{\sqrt{2}\omega^2}\right)} \right] \mathcal G_- - \left[1 - \cosh{\left( \frac{z\, \mathcal C_+ }{\sqrt{2}\omega^2}\right)} \right] \mathcal G_+  + \omega^6 \left[ \cosh{\left( \frac{z\, \mathcal C_+ }{\sqrt{2}\omega^2}\right)}  - \cosh{\left( \frac{z\, \mathcal C_- }{\sqrt{2}\omega^2}\right)}  \right]\right) \sigma^2 \bar B_x \bar B_z}{2\,\mathcal F \left(\omega^4 +\sigma^2 \bar B^2- \omega^2 m_\phi^2 \right) } A_z(0, \omega)\, + \nonumber\\
\frac{\sigma\bar B_x }{2 \,\omega^2\,\mathcal F } \left[2 \omega^6 \left[ \cosh{\left( \frac{z\, \mathcal C_- }{\sqrt{2}\omega^2}\right)} - \cosh{\left( \frac{z\, \mathcal C_+ }{\sqrt{2}\omega^2}\right)} \right] - \frac{\sigma^2 \bar B_z^2}{\omega^4 +\sigma^2 \bar B^2- \omega^2 m_\phi^2 } \left( 2\omega^6 \cosh{\left( \frac{z\, \mathcal C_- }{\sqrt{2}\omega^2}\right)} - \mathcal G_- \left[ 1 -\cosh{\left( \frac{z\, \mathcal C_- }{\sqrt{2}\omega^2}\right)}  \right] \right.\right. \nonumber\\ + \left.\left.  \mathcal G_+ - (2\omega^6 + \mathcal G_+) \cosh{\left( \frac{z\, \mathcal C_+ }{\sqrt{2}\omega^2}\right)}  \right) \right] \phi(0, \omega) + \frac{\left[\sinh{\left( \frac{z\, \mathcal C_- }{\sqrt{2}\omega^2}\right)} \mathcal C_+ - \sinh{\left( \frac{z\, \mathcal C_+ }{\sqrt{2}\omega^2}\right)} \mathcal C_- \right] 2\sqrt{2}\, \sigma\, \omega^6 \bar B_x}{2\mathcal F \mathcal C_- \mathcal C_+ } \partial_z \phi(0, \omega), \nonumber
\end{gather*}
\begin{gather*}
A_y(z, \omega)= \frac{\left(\left[\cos(z\omega) - \cosh{\left( \frac{z\, \mathcal C_- }{\sqrt{2}\omega^2}\right)} \right]\mathcal G_- - \left[ \cos(z\omega) - \cosh{\left( \frac{z\, \mathcal C_+ }{\sqrt{2}\omega^2}\right)} \mathcal G_+ \right] \right) \bar B_x \bar B_y}{2\mathcal F\bar B_T^2} A_x(0, \omega)\, - \nonumber \\  \frac{\left(\left[\sinh{\left( \frac{z\, \mathcal C_- }{\sqrt{2}\omega^2}\right)} \mathcal G_- \mathcal C_+ - \sinh{\left( \frac{z\, \mathcal C_+ }{\sqrt{2}\omega^2}\right)} \mathcal G_+ \mathcal C_- \right] \sqrt{2}\omega^3  + 2\mathcal F \mathcal C_-\mathcal C_+ \sin(z\omega) \right) \bar B_x \bar B_y }{2\,\omega \,\mathcal C_-\, \mathcal C_+\,\mathcal F \bar B_T^2} \partial_z A_x(0, \omega)\, + \nonumber \\  \frac{\left[\cosh{\left( \frac{z\, \mathcal C_+ }{\sqrt{2}\omega^2}\right)} \mathcal G_+ - \cosh{\left( \frac{z\, \mathcal C_- }{\sqrt{2}\omega^2}\right)} \mathcal G_- \right] \bar B_y^2 + 2 \cos(z\omega) \mathcal F\bar B_x^2}{2\mathcal F \bar B_T^2}  A_y(0, \omega)\, + \nonumber\\
\frac{\left[\sinh{\left( \frac{z\, \mathcal C_+ }{\sqrt{2}\omega^2}\right)} \mathcal G_+ \mathcal C_- - \sinh{\left( \frac{z\, \mathcal C_- }{\sqrt{2}\omega^2}\right)} \mathcal G_- \mathcal C_+ \right] \sqrt{2}\omega^3 \bar B_y^2 + 2\mathcal F \mathcal C_-\mathcal C_+ \sin(z\omega) \bar B_x^2}{2\,\omega \,\mathcal C_-\, \mathcal C_+\,\mathcal F \bar B_T^2} \partial_z A_y(0, \omega) + \nonumber \\
\frac{\left( \left[1 - \cosh{\left( \frac{z\, \mathcal C_- }{\sqrt{2}\omega^2}\right)} \right] \mathcal G_- - \left[1 - \cosh{\left( \frac{z\, \mathcal C_+ }{\sqrt{2}\omega^2}\right)} \right] \mathcal G_+  + \omega^6 \left[ \cosh{\left( \frac{z\, \mathcal C_+ }{\sqrt{2}\omega^2}\right)}  - \cosh{\left( \frac{z\, \mathcal C_- }{\sqrt{2}\omega^2}\right)}  \right]\right) \sigma^2 \bar B_y \bar B_z}{2\,\mathcal F \left(\omega^4 +\sigma^2 \bar B^2- \omega^2 m_\phi^2 \right) } A_z(0, \omega)\, + \nonumber\\
\frac{\sigma\bar B_y }{2 \,\omega^2\,\mathcal F } \left[2 \omega^6 \left[ \cosh{\left( \frac{z\, \mathcal C_- }{\sqrt{2}\omega^2}\right)} - \cosh{\left( \frac{z\, \mathcal C_+ }{\sqrt{2}\omega^2}\right)} \right] - \frac{\sigma^2 \bar B_z^2}{\omega^4 +\sigma^2 \bar B^2- \omega^2 m_\phi^2 } \left( 2\omega^6 \cosh{\left( \frac{z\, \mathcal C_- }{\sqrt{2}\omega^2}\right)} - \mathcal G_- \left[ 1 -\cosh{\left( \frac{z\, \mathcal C_- }{\sqrt{2}\omega^2}\right)}  \right] \right.\right. \nonumber\\ + \left.\left.  \mathcal G_+ - (2\omega^6 + \mathcal G_+) \cosh{\left( \frac{z\, \mathcal C_+ }{\sqrt{2}\omega^2}\right)}  \right) \right] \phi(0, \omega) + \frac{\left[\sinh{\left( \frac{z\, \mathcal C_- }{\sqrt{2}\omega^2}\right)} \mathcal C_+ - \sinh{\left( \frac{z\, \mathcal C_+ }{\sqrt{2}\omega^2}\right)} \mathcal C_- \right] 2\sqrt{2}\, \sigma\, \omega^6 \bar B_y}{2\mathcal F \mathcal C_- \mathcal C_+ } \partial_z \phi(0, \omega), \nonumber
\end{gather*}
\begin{gather*}
A_z(z, \omega)= \frac{\left(\left[ \cosh{\left( \frac{z\, \mathcal C_- }{\sqrt{2}\omega^2}\right)} -  \cosh{\left( \frac{z\, \mathcal C_+ }{\sqrt{2}\omega^2}\right)}  \right] \right) \mathcal G_+ \mathcal G_- \bar B_x \bar B_z}{4\,\omega^6\,\mathcal F\bar B_T^2} A_x(0, \omega)\, + \frac{\left(\left[ \sinh{\left( \frac{z\, \mathcal C_- }{\sqrt{2}\omega^2}\right)}\mathcal C_+ -  \sinh{\left( \frac{z\, \mathcal C_+ }{\sqrt{2}\omega^2}\right)} \mathcal C_- \right] \right) \mathcal G_+ \mathcal G_- \bar B_x \bar B_z}{2\sqrt{2}\,\omega^4\,\mathcal F\,\mathcal C_-\,\mathcal C_+\bar B_T^2} \partial_z A_x(0, \omega)\,  \nonumber \\ + \frac{\left(\left[ \cosh{\left( \frac{z\, \mathcal C_- }{\sqrt{2}\omega^2}\right)} -  \cosh{\left( \frac{z\, \mathcal C_+ }{\sqrt{2}\omega^2}\right)}  \right] \right) \mathcal G_+ \mathcal G_- \bar B_y \bar B_z}{4\,\omega^6\,\mathcal F\bar B_T^2}\,  A_y(0, \omega)\, +  \frac{\left(\left[ \sinh{\left( \frac{z\, \mathcal C_- }{\sqrt{2}\omega^2}\right)}\mathcal C_+ -  \sinh{\left( \frac{z\, \mathcal C_+ }{\sqrt{2}\omega^2}\right)}\mathcal C_-  \right] \right) \mathcal G_+ \mathcal G_- \bar B_y \bar B_z}{2\sqrt{2}\,\omega^4\,\mathcal F\,\mathcal C_-\,\mathcal C_+\bar B_T^2} \partial_z A_y(0, \omega) + \nonumber \\
\left[2\omega^6 \mathcal G_+ \left( \omega^4 +\sigma^2 \left[ \bar B_T^2 + \bar B_z^2 \cosh{\left( \frac{z\, \mathcal C_- }{\sqrt{2}\omega^2}\right)} \right] -\omega^2 m_\phi^2 \right) - \mathcal G_- \left( 2\omega^{10} + \sigma^2 \left[ 2\omega^6 \bar B_T^2 + \left(  \cosh{\left( \frac{z\, \mathcal C_+ }{\sqrt{2}\omega^2}\right)} (2\omega^6 + \mathcal G_+) -  \right. \right. \right.\right. \nonumber \\ \left. \left.\left.\left. \cosh{\left( \frac{z\, \mathcal C_- }{\sqrt{2}\omega^2}\right)} \mathcal G_+ \right) \bar B_z^2 \right]  - 2\,\omega^8 m_\phi^2 \right) \right] \left[ 4\,\omega^6 \mathcal F \left(\omega^4 +\sigma^2 \bar B^2- \omega^2 m_\phi^2 \right) \right]^{-1} A_z(0, \omega)\, + \nonumber\\
\sigma \bar B_z\left[2\omega^6 \left[1- \cosh{\left( \frac{z\, \mathcal C_- }{\sqrt{2}\omega^2}\right)} \right] \mathcal G_+\left(\omega^4 +\sigma^2 \bar B_T^2 -\omega^2 m_\phi^2 \right)- \mathcal G_- \left( 2\omega^{10}\left(1 - \cosh{\left( \frac{z\, \mathcal C_+ }{\sqrt{2}\omega^2}\right)} \right) + \sigma^2 \left[ 2\omega^6 \bar B_T^2 \left( 1-   \right. \right. \right.\right. \nonumber \\ \left. \left.\left.\left. \cosh{\left( \frac{z\, \mathcal C_+ }{\sqrt{2}\omega^2}\right)} \right)+ \left( \cosh{\left( \frac{z\, \mathcal C_+ }{\sqrt{2}\omega^2}\right)} -  \cosh{\left( \frac{z\, \mathcal C_- }{\sqrt{2}\omega^2}\right)} \right) \mathcal G_+ \bar B_z^2 \right] - 2\,\omega^8 m_\phi^2 \left( 1-  \cosh{\left( \frac{z\, \mathcal C_+ }{\sqrt{2}\omega^2}\right)}  \right) \right)\right] \times\nonumber \\ \left[ 4\,\omega^8 \mathcal F \left(\omega^4 +\sigma^2 \bar B^2- \omega^2 m_\phi^2 \right) \right]^{-1} \phi(0, \omega)\, + \frac{\left[\sinh{\left( \frac{z\, \mathcal C_+ }{\sqrt{2}\omega^2}\right)} \mathcal C_- \mathcal G_- - \sinh{\left( \frac{z\, \mathcal C_- }{\sqrt{2}\omega^2}\right)} \mathcal C_+ \mathcal G_+ \right] \sqrt{2}\, \sigma\, \bar B_z}{2\mathcal F \mathcal C_- \mathcal C_+ } \partial_z \phi(0, \omega), \nonumber
\end{gather*}
\begin{equation}\label{sys-1}
\begin{gathered}
\phi(z, \omega)= \frac{\left(\left[ \cosh{\left( \frac{z\, \mathcal C_+ }{\sqrt{2}\omega^2}\right)} -  \cosh{\left( \frac{z\, \mathcal C_- }{\sqrt{2}\omega^2}\right)}  \right] \right) \mathcal G_+ \mathcal G_- \bar B_x}{4\,\sigma\,\omega^4\,\mathcal F\bar B_T^2} A_x(0, \omega)\, - \frac{\left(\left[ \sinh{\left( \frac{z\, \mathcal C_- }{\sqrt{2}\omega^2}\right)}\mathcal C_+ -  \sinh{\left( \frac{z\, \mathcal C_+ }{\sqrt{2}\omega^2}\right)} \mathcal C_- \right] \right) \mathcal G_+ \mathcal G_- \bar B_x }{2\sqrt{2}\,\sigma\,\omega^2\,\mathcal F\,\mathcal C_-\,\mathcal C_+\bar B_T^2} \partial_z A_x(0, \omega)\,  -  \\ \frac{\left(\left[ \cosh{\left( \frac{z\, \mathcal C_- }{\sqrt{2}\omega^2}\right)} -  \cosh{\left( \frac{z\, \mathcal C_+ }{\sqrt{2}\omega^2}\right)}  \right] \right) \mathcal G_+ \mathcal G_- \bar B_y }{4\,\sigma\,\omega^4\,\mathcal F\bar B_T^2}\,  A_y(0, \omega)\, -  \frac{\left(\left[ \sinh{\left( \frac{z\, \mathcal C_- }{\sqrt{2}\omega^2}\right)}\mathcal C_+ -  \sinh{\left( \frac{z\, \mathcal C_+ }{\sqrt{2}\omega^2}\right)}\mathcal C_-  \right] \right) \mathcal G_+ \mathcal G_- \bar B_y}{2\sqrt{2}\,\sigma\,\omega^2\,\mathcal F\,\mathcal C_-\,\mathcal C_+\bar B_T^2} \partial_z A_y(0, \omega)\, +\\
 \frac{\sigma \bar B_z}{4\,\omega^4 \mathcal F \left(\omega^4 +\sigma^2 \bar B^2- \omega^2 m_\phi^2 \right)} \left[ 2\omega^6\mathcal G_- \left[ \cosh{\left( \frac{z\, \mathcal C_+ }{\sqrt{2}\omega^2}\right)} -1\right] + \left[ 2\omega^6  + \mathcal G_- \cosh{\left( \frac{z\, \mathcal C_+ }{\sqrt{2}\omega^2}\right)} - (2\omega^6 + \mathcal G_-) \cosh{\left( \frac{z\, \mathcal C_- }{\sqrt{2}\omega^2}\right)} \right] \mathcal G_+ \right]A_z(0, \omega)\\
 + \left[2\omega^6 \left[\sigma^2 \bar B_z^2 + \cosh{\left( \frac{z\, \mathcal C_- }{\sqrt{2}\omega^2}\right)} \right] \mathcal G_+\left(\omega^4 +\sigma^2 \bar B_T^2 -\omega^2 m_\phi^2 \right)- \mathcal G_- \left( 2\omega^{10} \cosh{\left( \frac{z\, \mathcal C_+ }{\sqrt{2}\omega^2}\right)} + \sigma^2 \left[ 2\omega^6 \bar B_T^2  \right. \right.\right.  \\ \left. \left.\left. \cosh{\left( \frac{z\, \mathcal C_+ }{\sqrt{2}\omega^2}\right)} + \left[ 2\omega^6 +\left( \cosh{\left( \frac{z\, \mathcal C_- }{\sqrt{2}\omega^2}\right)} -  \cosh{\left( \frac{z\, \mathcal C_+ }{\sqrt{2}\omega^2}\right)} \right) \mathcal G_+ \right]\bar B_z^2 \right] - 2\,\omega^8 m_\phi^2   \cosh{\left( \frac{z\, \mathcal C_+ }{\sqrt{2}\omega^2}\right)} \right)\right] \times \\ \left[ 4\,\omega^6 \mathcal F \left(\omega^4 +\sigma^2 \bar B^2- \omega^2 m_\phi^2 \right) \right]^{-1} \phi(0, \omega)\, + \frac{\left[\sinh{\left( \frac{z\, \mathcal C_- }{\sqrt{2}\omega^2}\right)} \mathcal C_+ \mathcal G_+ - \sinh{\left( \frac{z\, \mathcal C_+ }{\sqrt{2}\omega^2}\right)} \mathcal C_- \mathcal G_- \right] \sqrt{2}\, \omega^2}{2\mathcal F \mathcal C_- \mathcal C_+ } \partial_z \phi(0, \omega).
\end{gathered}
\end{equation}

The condition that Det[$P]\neq 0 $ is a sufficient condition that $M$ is diagonalizable. Since in this work we calculate the matrix exponential for only diagonalizable matrices, we keep in mind that we must work with values of the parameter space $\{\omega, \bar B_i, \Pi_{\mu\nu}, m_\phi, g_{\phi\gamma}\}$ that satisfy the condition Det[$P]\neq 0 $.  This is usually satisfied when $\mathcal C_\pm \neq 0, \mathcal G_\pm \neq 0, \mathcal F\neq 0$ and $\omega^4 +\sigma^2 \bar B^2- \omega^2 m_\phi^2 \neq 0$. Obviously, $\mathcal F$ is never zero for $\omega\neq 0$. Also $\mathcal G_\pm=0$ has no solution if $\omega\neq 0$ is considered the independent variable as we have explicitly checked. One can also check that for $\omega^2= g_{\phi\gamma}^2 \bar B^2 +m_\phi^2$ we also have that $\mathcal C_+=0$ while $\mathcal C_-$ is never zero. So, the condition Det[$P]\neq 0 $ implies only that $\omega^2 \neq g_{\phi\gamma}^2 \bar B^2 +m_\phi^2$, namely the energy square must be different from the pseudoscalar particle effective mass square, $m_{\phi, \text{eff}}^{2}= g_{\phi\gamma}^2 \bar B^2 +m_\phi^2$. If we require to have propagating pseudoscalar particles, we must have that $\omega^2 > m_{\phi, \text{eff}}^{2}$. The condition $\omega^2 = m_{\phi, \text{eff}}^{2}$ means that pseudoscalar particles in a background magnetic field are at rest and do not propagate in space because they have zero momentum. Consequently, the condition Det[$P]\neq 0 $, implies that we are looking for solution to the equations of motion for propagating pseudoscalar particles only.

Another complementary consideration to be made is that since we are considering only dispersive phenomena of the electromagnetic radiation in a magnetized vacuum, the arguments of the hyperbolic trigonometric functions have to be purely imaginary. If we had these arguments to have also a real part, one can easily see that the solutions found would grow infinitely with the distance $z$. Since in the solutions \eqref{sys-1} appear $\mathcal C_\pm$ as arguments of the hyperbolic trigonometric functions, the conditions that the arguments must be purely imaginary implies that $\mathcal C_\pm$ must be imaginary. The latter conditions are satisfied when $-2\omega^6 -\sigma^2\omega^2\bar B_z^2 +\omega^4 m_\phi^2 \pm \mathcal F<0$ which only solution in terms of $\omega$ being the independent variable is $\omega^2 > g_{\phi\gamma}^2 \bar B^2 +m_\phi^2$. So we see that  dispersive phenomena of the electromagnetic radiation in a magnetized vacuum imply that $\omega^2 > g_{\phi\gamma}^2 \bar B^2 +m_\phi^2$, namely propagating fields as discussed above. In the case when $\omega^2 < g_{\phi\gamma}^2 \bar B^2 +m_\phi^2$ one can check that the electromagnetic and pseudoscalar fields do not propagate in space and they become evanescent or get damped with the propagation distance.

\section{Solution of equations of motion in a magnetized plasma/gas}
\label{sec:4}

In the previous section, we found the solutions of the equation of motion in the case when the mixing of electromagnetic waves with the pseudoscalar field happens in the presence of an external magnetic field only. In this section, we want to focus on the case when in addition to the external magnetic field there is present also a medium. However, the solutions of the equations of motion in the general case when both media and an external magnetic field coexist are very difficult to find because it is extremely hard to calculate the matrix exponential in \eqref{sol-eq-mot}. In this section, we concentrate on the case when the external magnetic field is completely perpendicular to the direction of propagation of the electromagnetic field/pseudoscalar field that we choose for simplicity to be $\bar{\bs B} =(\bar B_x, 0, 0)$. Besides, we need to specify the type of medium where the fields propagate in. Usually, the mixing of photons with pseudoscalar particles takes place in a vacuum (simplest case) or a gas or a plasma. Let us consider the case when the mixing occurs in a magnetized plasma of electrons and protons or other heavy ionized nuclei. The reason for this choice is that the elements of the photon polarization tensor are explicitly known in the case of a cold collisionless plasma, see Ref. \cite{Ejlli:2018ucq} for details. Also, the solution that we find below is valid in the case of a gas or another type of media in which the photon polarization tensor has the same non zero elements. For a cold collisionless plasma the only non zero elements of the photon poalarization tensor are $\Pi_{xx, yy, yz,0z}$, see \cite{Ejlli:2018ucq} for details. In this case in order to find the eigenvalues and eigenvectors, we first must solve the characteristic polynomial equation of the eigenvalues 
\begin{equation}\label{pol-1}
 \left[ \left(\lambda \omega - i\Pi_{0z} \right) \left(\lambda^2 +\omega^2 - \Pi_{yy} \right) - i\Pi_{0y} \Pi_{yz}\right] \left[ \left(\lambda^2 +\omega^2 - m_\phi^2 \right) \left(\lambda^2 +\omega^2 - \Pi_{xx} \right) -\omega^2 g_{\phi\gamma}^2  \bar B_x^2 \right] =0.
\end{equation}

The characteristic polynomial equation \eqref{pol-1} is of degree of seven and it has seven distinct or partially repetitive roots. However the roots of the characteristic polynomial equations are usually very complex to find and it may be useful to make some approximation at this point. The approximation that we make is to neglect the term $i\Pi_{0y} \Pi_{yz}$ with respect to the term $i\Pi_{0z} \Pi_{yy}$ in the first part of equation \eqref{pol-1}. This is usually very accurate since in the case of a plasma the element $\Pi_{0y}= k \Pi_{zy}/\omega\simeq \Pi_{zy}$ (by using the continuity and gauge invariance conditions) is the term related to the Faraday effect which mixes the $A_y$ and $A_z$ states and which is usually a small term for ordinary laboratory magnetic field strengths and high frequency electromagnetic waves. This can be seen by using explicitly the expressions of the elements of the photon polarization tensor as shown in Eq. (7) of Ref. \cite{Ejlli:2018ucq} in the case of perpendicular propagation with respect to the external magnetic field that corresponds to $\Theta=0$ and $\Phi=0$, see Eq. (7) of Ref. \cite{Ejlli:2018ucq} for details:
\begin{equation}\label{eleph}
\Pi^{yy}(\omega)=\Pi^{zz}(\omega)=\frac{\omega_\text{pl}^2 \omega^2}{\omega^2-\omega_c^2}, \qquad \Pi^{yz}(\omega)=\Pi^{zy*}(\omega)= -i \frac{\omega_\text{pl}^2\omega_c\omega}{\omega^2-\omega_c^2}, \qquad \omega\neq \omega_c,
\end{equation}
where $\omega_c = e \bar B/m_e\simeq 1.76\times 10^7(\bar B_x/\text{G})$ (rad/s) is the cyclotron frequency and $m_e$ is the electron mass. By using the the continuity and gauge invariance conditions for $k\simeq \omega$, the condition $i\Pi_{0y} \Pi_{yz}\ll i\Pi_{0z} \Pi_{yy}$
translates into $\Pi^{yy}\Pi^{zz}\gg |\Pi^{yz}|^2$. The latter condition implies by using \eqref{eleph} that $\omega^2\gg \omega_c^2$ which in numbers translates into $\nu=\omega/2\pi\gg 2.8\times 10^{10} (\bar B_x/\text{T})$ (Hz), with $\nu$ being the incident electromagnetic wave frequency. Since most pseudoscalar particle detectors operate in the frequency regime $10^{14}$ (Hz)-$10^{18}$ (Hz) and for magnetic field strengths of few Teslas (T), we can easily see that our assumption to neglect the term $i\Pi_{0y} \Pi_{yz}$ with respect to the term $i\Pi_{0z} \Pi_{yy}$ is justified. Clearly, this approximation is also satisfied in the case of cosmological magnetic fields.

Under the approximation discussed above\footnote{Here we are slightly deviating from our initial assertion of exact solution of Eqs. \eqref{system-1} in the case of a plasma/gas since we made the justified approximation to neglect the term $i\Pi_{0y} \Pi_{yz}$ with respect to the term $i\Pi_{0z} \Pi_{yy}$ in \eqref{pol-1} before finding the solutions \eqref{sys-2}. Clearly, the exact solution of \eqref{system-1} is still possible as we have explicitly checked without using the above approximation but it is quite cumbersome. }, the eigenvalues obtained from the solution of equation \eqref{pol-1} and their respective eigenvectors are given by
\begin{equation}\label{Eigen-2}
\begin{gathered}
\lambda_1= i\Pi_{0z}/\omega, \quad \lambda_2 = -\mathcal A_-/ \sqrt{2}, \quad \lambda_3 = \mathcal A_-/ \sqrt{2}, \quad \lambda_4 = -\mathcal A_+/ \sqrt{2}, \quad \lambda_5 = \mathcal A_+/ \sqrt{2}, \quad \lambda_6 = - \sqrt{\Pi_{yy} -\omega^2}, \quad \lambda_7 =  \sqrt{\Pi_{yy} -\omega^2} \\
v_1= \left[0, 0, \frac{\omega^2 \Pi_{yz}}{\omega^4 -\Pi_{0z}^2 -\omega^2 \Pi_{yy}}, \frac{i\Pi_{0z} \Pi_{yz}}{\omega^4 -\Pi_{0z}^2 -\omega^2 \Pi_{yy}}, 1, 0, 0\right]^\text{T}, \\ v_2= \left[\frac{\sqrt{2}\sigma \bar B_x \mathcal A_-}{-2\sigma^2 \bar B_x^2 + (\omega^2 - m_\phi^2)\left( \Pi_{xx} - m_\phi^2 + \mathcal S\right)}, \frac{ \left( \Pi_{xx} - m_\phi^2 -\mathcal S \right)}{2\sigma \bar B_x}, 0, 0, 0, -\frac{\sqrt{2}}{\mathcal A_-}, 1\right]^\text{T}, \\
v_3= \left[-\frac{\sqrt{2}\sigma \bar B_x \mathcal A_-}{-2\sigma^2 \bar B_x^2 + (\omega^2 - m_\phi^2)\left( \Pi_{xx} - m_\phi^2 + \mathcal S\right)}, \frac{ \left( \Pi_{xx} - m_\phi^2 -\mathcal S \right)}{2\sigma \bar B_x}, 0, 0, 0, \frac{\sqrt{2}}{\mathcal A_-}, 1\right]^\text{T},\\
v_4= \left[-\frac{\sqrt{2}\sigma \bar B_x \mathcal A_+}{2\sigma^2 \bar B_x^2 + (\omega^2 - m_\phi^2)\left(- \Pi_{xx} + m_\phi^2 + \mathcal S\right)}, \frac{ \left( \Pi_{xx} - m_\phi^2 + \mathcal S \right)}{2\sigma \bar B_x}, 0, 0, 0, -\frac{\sqrt{2}}{\mathcal A_+}, 1\right]^\text{T},\\
v_5= \left[\frac{\sqrt{2}\sigma \bar B_x \mathcal A_+}{2\sigma^2 \bar B_x^2 + (\omega^2 - m_\phi^2)\left(- \Pi_{xx} + m_\phi^2 + \mathcal S\right)}, \frac{ \left( \Pi_{xx} - m_\phi^2 + \mathcal S \right)}{2\sigma \bar B_x}, 0, 0, 0, \frac{\sqrt{2}}{\mathcal A_+}, 1\right]^\text{T},\\
v_6= \left[0, 0, -\frac{1}{\sqrt{\Pi_{yy}-\omega^2}}, 1, 0, 0, 0 \right]^\text{T},\quad v_7= \left[0, 0, \frac{1}{\sqrt{\Pi_{yy}-\omega^2}}, 1, 0, 0, 0 \right]^\text{T},
\end{gathered}
\end{equation}
where we have defined
\begin{equation}\label{A-S-def}
\mathcal S \equiv \sqrt{-4\sigma^2 \bar B_x^2 + (m_\phi^2 - \Pi_{xx})^2}, \qquad \mathcal A_\pm \equiv \sqrt{-2\omega^2 + m_\phi^2 + \Pi_{xx} \pm \mathcal S}.
\end{equation}

Now that we have found the eigenvalues and the corresponding eigenvectors, the solution of equations of motion are given by taking the matrix exponential in \eqref{system-1} and after many lengthy calculations we get
\begin{equation}\label{sys-2}
\begin{gathered}
A_x(z, \omega)= \left( \frac{1}{2}\left[ \cosh{\left( \frac{z\, \mathcal A_- }{\sqrt{2}}\right)} +  \cosh{\left( \frac{z\, \mathcal A_+ }{\sqrt{2}}\right)}  \right]  + \frac{m_\phi^2 - \Pi_{xx}}{2 \mathcal S} \left[ \cosh{\left( \frac{z\, \mathcal A_- }{\sqrt{2}}\right)} -  \cosh{\left( \frac{z\, \mathcal A_+ }{\sqrt{2}}\right)}  \right] \right) A_x(0, \omega)\: +  \\
\frac{1}{\sqrt{2} \mathcal S} \left(  \frac{\sinh\left({\frac{z \mathcal A_-}{\sqrt{2}}}\right)\left[ m_\phi^2 - \Pi_{xx} + \mathcal S \right]}{\mathcal A_-}     - \frac{\sinh\left({\frac{z \mathcal A_+}{\sqrt{2}}}\right)\left[ m_\phi^2 - \Pi_{xx} - \mathcal S \right]}{\mathcal A_+}\right) \partial_z A_x(0, \omega) + \frac{\sigma \bar B_x}{\mathcal S}\left[ \cosh{\left( \frac{z\, \mathcal A_- }{\sqrt{2}}\right)} -  \cosh{\left( \frac{z\, \mathcal A_+ }{\sqrt{2}}\right)}  \right] \phi(0, \omega) \\
+  \frac{\sqrt{2}\sigma \bar B_x}{\mathcal S}\left[ \frac{\sinh{\left( \frac{z\, \mathcal A_- }{\sqrt{2}}\right)}}{\mathcal A_-} -  \frac{\sinh{\left( \frac{z\, \mathcal A_+ }{\sqrt{2}}\right)}}{\mathcal A_+}  \right] \partial_z \phi(0, \omega), \\
A_y(z, \omega) = \cosh{\left(z \sqrt{\Pi_{yy} -\omega^2} \right)} A_y(0, \omega) +  \frac{\sinh{\left(z \sqrt{\Pi_{yy} -\omega^2} \right)} }{\sqrt{\Pi_{yy} -\omega^2} } \partial_z A_y(0, \omega) + \frac{\omega^2 \Pi_{yz}}{2} \left( \frac{2 e^{\frac{i z \Pi_{0z}}{\omega}}}{\omega^4 - \Pi_{0z}^2 -\omega^2 \Pi_{yy}}  \right. \\  \left. - \frac{e^{- z \sqrt{\Pi_{yy}-\omega^2}}}{\omega^4 - i \Pi_{0z}\sqrt{\omega^2 \Pi_{yy} -\omega^4} -\omega^2 \Pi_{yy}}  - \frac{e^{ z \sqrt{\Pi_{yy}-\omega^2}}}{\omega^4 + i \Pi_{0z}\sqrt{\omega^2 \Pi_{yy} -\omega^4} -\omega^2 \Pi_{yy}}  \right) A_z(0, \omega),\\
A_z(z, \omega) = e^{iz \Pi_{0z}/\omega} A_z(0, \omega), \\
\phi(z, \omega) = \frac{\sigma \bar B_x}{\mathcal S}\left[ \cosh{\left( \frac{z\, \mathcal A_+ }{\sqrt{2}}\right)} -  \cosh{\left( \frac{z\, \mathcal A_- }{\sqrt{2}}\right)}  \right] A_x(0, \omega) - \frac{\sqrt{2}\sigma \bar B_x}{\mathcal S}\left[ \frac{\sinh{\left( \frac{z\, \mathcal A_- }{\sqrt{2}}\right)}}{\mathcal A_-} -  \frac{\sinh{\left( \frac{z\, \mathcal A_+ }{\sqrt{2}}\right)}}{\mathcal A_+}  \right] \partial_z A_x (0, \omega) + \\
 \left( \frac{1}{2}\left[ \cosh{\left( \frac{z\, \mathcal A_- }{\sqrt{2}}\right)} +  \cosh{\left( \frac{z\, \mathcal A_+ }{\sqrt{2}}\right)}  \right]  - \frac{m_\phi^2 - \Pi_{xx}}{2 \mathcal S} \left[ \cosh{\left( \frac{z\, \mathcal A_- }{\sqrt{2}}\right)} -  \cosh{\left( \frac{z\, \mathcal A_+ }{\sqrt{2}}\right)}  \right] \right) \phi(0, \omega)\: + \\
 \frac{1}{\sqrt{2} \mathcal S} \left(  \frac{\sinh\left({\frac{z \mathcal A_+}{\sqrt{2}}}\right)\left[ m_\phi^2 - \Pi_{xx} + \mathcal S \right]}{\mathcal A_+}     - \frac{\sinh\left({\frac{z \mathcal A_-}{\sqrt{2}}}\right)\left[ m_\phi^2 - \Pi_{xx} - \mathcal S \right]}{\mathcal A_-}\right) \partial_z \phi(0, \omega).
 \end{gathered}
 \end{equation}

As we did in Sec. \ref{sec:3}, the condition for the matrix $M$ to be diagonalizable in the case studied in this section is that the determinant of the matrix $P$ formed with the eigenvectors in \eqref{Eigen-2}, Det[$P]\neq 0 $, which is given by
\begin{equation}
\text{Det} [P] = \frac{16\, (2\sigma \bar B_x + m_\phi^2 - \Pi_{xx})\, (2\sigma \bar B_x - m_\phi^2 + \Pi_{xx})}{\sigma^2 \bar B_x^2\,\mathcal A_-\mathcal A_+ \sqrt{\Pi_{yy} -\omega^2} } \neq 0
\end{equation}
The solutions of Det[$P$]=0 in terms of the energy $\omega$ being the independent variable are $\omega = \pm (i/2 g_{\phi\gamma} \bar B_x) (m_\phi^2 - \Pi_{xx})$. However, since the energy is a real variable, we essentially have that the condition Det[$P] \neq 0$ is always true for $\omega\neq 0$ in the case when $\Pi_{xx}$ is a real quantity. Because of the fact that we are considering the case of only dispersive phenomena related to the photon polarization tensor, we have indeed that $\Pi_{xx}$ is real because $\Pi_{ij}=\Pi_{ji}^*$, namely a hermitian tensor.

Also we must have that $\mathcal A_\pm\neq 0$ and $\sqrt{\Pi_{yy}-\omega^2} \neq 0$ for $\omega\neq 0$ in order that  Det[$P$] is a finite quantity. The condition $\sqrt{\Pi_{yy}-\omega^2} \neq 0$ is also satisfied if we want to have a propagating photon mode along the $y$ direction where $\omega^2 > \Pi_{yy}$, namely the photon energy square must be bigger than the effective photon mass square in the medium. Clearly, in the case when $\omega^2 < \Pi_{yy}$, it is known that in the case of a plasma, the electromagnetic radiation does not propagate and it gets damped.
The conditions $\mathcal A_\pm \neq 0$ are satisfied when
\begin{equation}\label{om-cond}
\omega^2 \neq \frac{m_{\text{eff},\,\phi}^2 +\Pi_{xx}  \pm \sqrt{\left( m_{\text{eff},\,\phi}^2  -\Pi_{xx} \right)^2 + 4g_{\phi\gamma}^2 \bar B_x^2\, \Pi_{xx}}}{2},
\end{equation}
where $m_{\text{eff},\,\phi}^2=m_\phi^2 + g_{\phi\gamma}^2 \bar B_x^2$. In the case when $\{m_{\text{eff}, \phi}^4, \Pi_{xx}^2 \}\gg 4g_{\phi\gamma}^2 \bar B_x^2\, \Pi_{xx}$, we would have that $\omega^2 \neq m_{\text{eff},\,\phi}^2$ and $\omega^2 \neq \Pi_{xx}$. These conditions clearly have a physical meaning, where the first is to have an energy different from the effective pseudoscalar particle mass and the second the energy of the fields must also be different from effective mass of the photon in media. We may note that in the case when $\Pi_{xx}=0$ we get exactly the same condition for the energy as we found in the case of propagation in a vacuum in Sec. \ref{sec:3}. In the case when $\{m_{\text{eff}, \phi}^4, \Pi_{xx}^2 \}\gg 4g_{\phi\gamma}^2 \bar B_x^2\, \Pi_{xx}$, clearly we must have $\omega^2 > m_{\text{eff},\,\phi}^2$ and $\omega^2 > \Pi_{xx}$ for propagating photons and pseudoscalar particles in space.
Under these conditions, similarly as we found in Sec. \ref{sec:3}, the matrix $M$ has seven distinct eigenvalues and consequently is diagonalizable. A sufficient (but not necessary) condition for a matrix to be diagonalizable is that all its eigenvalues must be different. 

As discussed in Sec. \ref{sec:3}, for dispersive phenomena of the propagating fields, we must have that both $\mathcal A_\pm$ must be purely imaginary\footnote{The solution of the inequality -$2\omega^2 + m_\phi^2 + \Pi_{xx} + \mathcal S<0$ is given by \eqref{om-cond-1}, while the other inequality $-2\omega^2 + m_\phi^2 + \Pi_{xx} - \mathcal S<0$ has several solutions. These solutions are: $\Pi_{xx}=0$ and $\omega\neq 0$, or $\Pi_{xx}>0$ and $\omega^2 > \frac{m_{\text{eff},\,\phi}^2 +\Pi_{xx}  - \sqrt{\left( m_{\text{eff},\,\phi}^2  -\Pi_{xx} \right)^2 + 4g_{\phi\gamma}^2 \bar B_x^2\, \Pi_{xx}}}{2}$, or $\Pi_{xx}<0$. The unique solution of both  -$2\omega^2 + m_\phi^2 + \Pi_{xx} \pm \mathcal S<0$ is given by \eqref{om-cond-1}.}. These conditions are both simultaneously satisfied when both -$2\omega^2 + m_\phi^2 + \Pi_{xx} \pm \mathcal S<0$, which solution is
\begin{equation}\label{om-cond-1}
\omega^2 > \frac{m_{\text{eff},\,\phi}^2 +\Pi_{xx}  + \sqrt{\left( m_{\text{eff},\,\phi}^2  -\Pi_{xx} \right)^2 + 4g_{\phi\gamma}^2 \bar B_x^2\, \Pi_{xx}}}{2}.
\end{equation}
The condition \eqref{om-cond-1} is a sufficient condition for propagating electromagnetic and pseudoscalar fields in a dispersive media.

It is quite useful also to see the behavior of solutions \eqref{sys-2} in the case when the arguments of the trigonometric functions are real instead of pure imaginary. Consider for simplicity the solution of $A_y(z, \omega)$ in \eqref{sys-2} in the case when $A_z(0, \omega)=0$ where 
\begin{equation}\label{Ay-sol}
A_y(z, \omega) = \cosh{\left(z \sqrt{\Pi_{yy} -\omega^2} \right)} A_y(0, \omega) +  \frac{\sinh{\left(z \sqrt{\Pi_{yy} -\omega^2} \right)} }{\sqrt{\Pi_{yy} -\omega^2} } \partial_z A_y(0, \omega).
\end{equation}
In the case when $\omega^2> \Pi_{yy}$, we have that the arguments of the hyperbolic functions are imaginary and consequently the solution \eqref{Ay-sol} is an oscillatory function which is a linear combination of sine and cosine functions. In the case when $\omega^2< \Pi_{yy}$ the arguments of the hyperbolic functions are real and it may seem that the solution \eqref{Ay-sol},  $A_y(z, \omega)\rightarrow\infty$ for $z \rightarrow \infty$. However, this is not the case. Suppose that in the region where is located the magnetic field we can write for a monochromatic wave $A_y(z, \omega)=A(k, \omega) e^{ikz}$ where $\partial_z A_y(z, \omega) = ik A_y(z, \omega)$. By comparing the latter expression with the derivative of $A_y$ with respect to $z$ in \eqref{Ay-sol}, after some calculations we find that $|k|=i\sqrt{\Pi_{yy} -\omega^2} $. In the case of propagating wave in the positive direction, we get from \eqref{Ay-sol} 
\begin{equation}\label{Ay-sol-1}
A_y(z, \omega) = \left[\cosh{\left(z \sqrt{\Pi_{yy} -\omega^2} \right)} - \sinh{\left(z \sqrt{\Pi_{yy} -\omega^2} \right)} \right]A_y(0, \omega) = \exp{\left(-z \sqrt{\Pi_{yy} -\omega^2} \right)}A_y(0, \omega).
\end{equation}
The solution in \eqref{Ay-sol-1} represents a damping or evanescent propagating electromagnetic wave. In the case of a magnetized plasma we have that $\omega^2< \Pi_{yy}= \omega_\text{pl}^2\omega^2/(\omega^2 - \omega_c^2)$ where for a magnetized plasma with a transverse magnetic field to the direction of propagation of the wave, $\Pi_{xx}=\omega_\text{pl}^2$, $\Pi_{yy}= \omega_\text{pl}^2\omega^2/(\omega^2 - \omega_c^2)$ \cite{Ejlli:2018ucq}. So, we can see that the solution \eqref{Ay-sol} gives a well known result in the case of propagation in a magnetized plasma, namely that for $\omega^2< \omega_\text{pl}^2 + \omega_c^2$ the electromagnetic wave state $A_y(z, \omega)$ becomes evanescent. This condition on the energy does not automatically imply that also the state $A_x(z, \omega)$ does not propagate in space.

Another thing that is worth to note about solutions \eqref{sys-2} is that $A_z(z, \omega)$ is only proportional to $A_z(0, \omega)$. This is a consequence of the fact that we have neglected the term $i\Pi_{0y} \Pi_{yz}$ when we solved the characteristic equation \eqref{pol-1}. If we kept also the term $i\Pi_{0y} \Pi_{yz}$ in \eqref{pol-1}, then $A_z(z, \omega)$ in \eqref{sys-1} would be also proportional to $A_y(0, \omega)$ whereas the the solutions for $A_{x}(z, \omega)$ and $\phi(z, \omega)$ in \eqref{sys-1} would be invariant. So, it would be an error to conclude that $A_z(z, \omega)=0$ for $A_z(0, \omega)=0$ since in this case one has also to consider the term proportional to $A_y(0, z)$ that has been neglected in our approximation.  This fact can also be seen from the fourth equation in \eqref{eq-A-ph-2} which must be satisfied by all fields in the region where the magnetic field is located.

\section{Transition efficiencies and longitudinal electric field}
\label{sec:5}

In Sec. \ref{sec:3} and Sec. \ref{sec:4} we found a solution of the equations of motion in the case when the mixing happens in an external magnetic field only with arbitrary direction with respect to the direction of propagation of fields and in the case of a transverse magnetic field with respect to the direction of propagation in a magnetized plasma. In this section we focus on discussing the solutions found in Sec. \ref{sec:3} and Sec. \ref{sec:4} and their implications and also compare these solutions with those found by using approximate methods.  

Let us first focus on the solutions found in Sec. \ref{sec:3} that we found in the case of propagation in a magnetized vacuum for arbitrary direction of the external magnetic field. It will be more convenient in what follows to slightly rewrite the functions $\mathcal C_\pm$ and $\mathcal F$ that appear in \eqref{def-0}
\begin{gather}
\mathcal C_\pm = i\sqrt{2}\, \omega^3 \left[ 1 - \frac{1}{2} \left( \frac{g_{\phi\gamma}\bar B_z}{\omega}\right)^2 - \frac{m_\phi^2}{2\,\omega^2} \mp \frac{\mathcal F}{2\,\omega^6} \right]^{1/2},\nonumber\\
\frac{\mathcal F}{2\,\omega^6}= \left[  \left( \frac{g_{\phi\gamma}\bar B_T}{\omega}\right)^2 + \frac{1}{4}  \left( \frac{g_{\phi\gamma}\bar B_z}{\omega}\right)^4 +\left( \frac{m_\phi}{2\,\omega}\right)^2 \left[ \left( \frac{m_\phi}{\omega}\right)^2  + 2\left( \frac{g_{\phi\gamma}\bar B_z}{\omega}\right)^2   \right]\right]^{1/2},
\end{gather}
where we considered $\omega>0$ for simplicity. Since we intend to apply our results in case of a constant and uniform magnetic field such as in a laboratory, let us consider for simplicity that we have initially a linearly polarized and monochromatic electromagnetic wave with polarization vector along the $\hat{\bs x}$ axis. In this case we have that only $A_x(0, \omega)$ and $\partial_z A_x(0, \omega)$ in \eqref{sys-1} are different from zero and
\begin{equation}
\begin{gathered}
A_x(z, \omega)= \frac{\left[\cosh{\left( \frac{z\, \mathcal C_+ }{\sqrt{2}\omega^2}\right)} \mathcal G_+ - \cosh{\left( \frac{z\, \mathcal C_- }{\sqrt{2}\omega^2}\right)} \mathcal G_- \right] \bar B_x^2 + 2 \cos(z\omega) \mathcal F\bar B_y^2}{2\mathcal F(\bar B_x^2 + \bar B_y^2)} A_x(0, \omega)\, + \\  \frac{\left[\sinh{\left( \frac{z\, \mathcal C_+ }{\sqrt{2}\omega^2}\right)} \mathcal G_+ \mathcal C_- - \sinh{\left( \frac{z\, \mathcal C_- }{\sqrt{2}\omega^2}\right)} \mathcal G_- \mathcal C_+ \right] \sqrt{2}\omega^3 \bar B_x^2 + 2\mathcal F \mathcal C_-\mathcal C_+ \sin(z\omega) \bar B_y^2}{2\,\omega \,\mathcal C_-\, \mathcal C_+\,\mathcal F (\bar B_x^2 + \bar B_y^2)} \partial_z A_x(0, \omega),\\
A_y(z, \omega)= \frac{\left(\left[\cos(z\omega) - \cosh{\left( \frac{z\, \mathcal C_- }{\sqrt{2}\omega^2}\right)} \right]\mathcal G_- - \left[ \cos(z\omega) - \cosh{\left( \frac{z\, \mathcal C_+ }{\sqrt{2}\omega^2}\right)} \mathcal G_+ \right] \right) \bar B_x \bar B_y}{2\mathcal F(\bar B_x^2 + \bar B_y^2)} A_x(0, \omega)\, -  \\  \frac{\left(\left[\sinh{\left( \frac{z\, \mathcal C_- }{\sqrt{2}\omega^2}\right)} \mathcal G_- \mathcal C_+ - \sinh{\left( \frac{z\, \mathcal C_+ }{\sqrt{2}\omega^2}\right)} \mathcal G_+ \mathcal C_- \right] \sqrt{2}\omega^3  + 2\mathcal F \mathcal C_-\mathcal C_+ \sin(z\omega) \right) \bar B_x \bar B_y }{2\,\omega \,\mathcal C_-\, \mathcal C_+\,\mathcal F (\bar B_x^2 + \bar B_y^2)} \partial_z A_x(0, \omega),  \\  
A_z(z, \omega)= \frac{\left(\left[ \cosh{\left( \frac{z\, \mathcal C_- }{\sqrt{2}\omega^2}\right)} -  \cosh{\left( \frac{z\, \mathcal C_+ }{\sqrt{2}\omega^2}\right)}  \right] \right) \mathcal G_+ \mathcal G_- \bar B_x \bar B_z}{4\,\omega^6\,\mathcal F(\bar B_x^2 + \bar B_y^2)} A_x(0, \omega)\, + \frac{\left(\left[ \sinh{\left( \frac{z\, \mathcal C_- }{\sqrt{2}\omega^2}\right)}\mathcal C_+ -  \sinh{\left( \frac{z\, \mathcal C_+ }{\sqrt{2}\omega^2}\right)} \mathcal C_- \right] \right) \mathcal G_+ \mathcal G_- \bar B_x \bar B_z}{2\sqrt{2}\,\omega^4\,\mathcal F\,\mathcal C_-\,\mathcal C_+(\bar B_x^2 + \bar B_y^2)} \partial_z A_x(0, \omega),\\
\phi(z, \omega)= \frac{\left(\left[ \cosh{\left( \frac{z\, \mathcal C_+ }{\sqrt{2}\omega^2}\right)} -  \cosh{\left( \frac{z\, \mathcal C_- }{\sqrt{2}\omega^2}\right)}  \right] \right) \mathcal G_+ \mathcal G_- \bar B_x}{4\,\sigma\,\omega^4\,\mathcal F(\bar B_x^2 + \bar B_y^2)} A_x(0, \omega)\, - \frac{\left(\left[ \sinh{\left( \frac{z\, \mathcal C_- }{\sqrt{2}\omega^2}\right)}\mathcal C_+ -  \sinh{\left( \frac{z\, \mathcal C_+ }{\sqrt{2}\omega^2}\right)} \mathcal C_- \right] \right) \mathcal G_+ \mathcal G_- \bar B_x }{2\sqrt{2}\,\sigma\,\omega^2\,\mathcal F\,\mathcal C_-\,\mathcal C_+(\bar B_x^2 + \bar B_y^2)} \partial_z A_x(0, \omega).
\end{gathered}\label{sys-3}
\end{equation}

The solutions in \eqref{sys-3} are exact and have been obtain in the case of propagation of electromagnetic waves in a magnetized vacuum only. Solutions \eqref{sys-3} can be further simplified if we assume that initially we have a linearly polarized monochromatic\footnote{The case when the wave is not monochromatic or a wave packet is much more complicated since it involves integration over $\omega$. } plane wave propagating in vacuum with constant amplitude in space. In this case we can replace $\partial_z A_x(0, \omega)=ik A_x(0, \omega)= i\omega A_x(0, \omega)$ where $k=\omega$ is the initial magnitude of the plane electromagnetic wave with wave-vector $\bs k= (0, 0, k)$ at $z=0$. Let $I_\gamma(0, \omega) \equiv |A_x(0, \omega)|^2$ be the intensity\footnote{Here with intensity we mean that associated to electromagnetic field vector-potential $A_i$ and not to the electric field $E_i$.} of the incident monochromatic plane wave at $z=0$ and the pseudoscalar field intensity at a distance $z$ from the source, $I_\phi(z, \omega) \equiv |\phi(z, \omega)|^2$. Then the efficiency (or probability) of transformation of photons into pseudoscalar particles as a function of $z$ and $\omega$ is given by
\begin{equation}\label{prob-1}
P_{\gamma\phi}(z, \omega) \equiv \frac{I_\phi(z, \omega)}{I_\gamma(0, \omega)} = \frac{\mathcal G_+^2 \mathcal G_-^2 \bar B_x^2}{8 \mathcal F^2 |\sigma|^2\omega^2 \bar B_T^4} \left| (\sqrt{2}\omega^3)^{-1}\left[ \cosh{\left( \frac{z\, \mathcal C_+ }{\sqrt{2}\omega^2}\right)} -  \cosh{\left( \frac{z\, \mathcal C_- }{\sqrt{2}\omega^2}\right)}  \right]  -i \left[ \frac{\sinh{\left( \frac{z\, \mathcal C_- }{\sqrt{2}\omega^2}\right)}}{\mathcal C_-} -  \frac{\sinh{\left( \frac{z\, \mathcal C_+ }{\sqrt{2}\omega^2}\right)}}{ \mathcal C_+} \right]  \right|^2.
\end{equation}
Expression \eqref{prob-1} is an exact expression that gives the efficiency or probability of transformation of photons into pseudoscalar particles in the presence of a constant magnetic field of arbitrary direction for an initial incident linearly polarized monochromatic wave. Expression \eqref{prob-1} generalizes the transition probability obtained in the literature that has been derived by using approximation methods. That expression is also valid in the case of an incident linearly polarized plane wave-like electromagnetic wave with a space varying amplitude that satisfies $|\partial_z A_x(0, \omega)|\ll |i k A_x(0, \omega)| $. Another important fact about \eqref{prob-1} and related expressions below, is that it is a bounded quantity or more precisely $P_{\gamma\phi}(z, \omega) \leq 1$. This is due to the fact that the pseudoscalar field intensity, is a quantity that is equal or less than the electromagnetic field intensity for the case of transformation of photons into pseudoscalar particles. This can be seen from \eqref{prob-1} where at the denominator is the initial photon intensity and consequently $P_{\gamma\phi}\leq 1$. One can use expression \eqref{prob-1} and the condition  $P_{\gamma\phi}(z, \omega) \leq 1$ to constraint the pseudscalar particle parameter space, for fixed values of the other parameters. This essentially means that if some of the parameters that enter \eqref{prob-1} are fixed, we cannot put arbitrary values for the remaining parameters that enter in \eqref{prob-1}. One has to adjust these parameters in order to have $P_{\gamma\phi}\leq 1$, thus constrain them.

Now let us calculate the transition efficiency in the case when matter is present and the external magnetic field is completely transverse with respect to the direction of propagation $\bar {\bs B}=(\bar B_x, 0, 0)$. By using the solution for the pseudoscalar field in \eqref{sys-2} for an incident linearly polarized monochromatic wave along the $\hat{\bs x}$ only, we get the following expression for the transition efficiency of photons into pseudoscalar particles
\begin{equation}\label{prob-2}
P_{\gamma\phi}(z, \omega) \equiv   \left|\frac{\sigma \bar B_x}{\mathcal S} \right|^2 \left|\left[ \cosh{\left( \frac{z\, \mathcal A_+ }{\sqrt{2}}\right)} -  \cosh{\left( \frac{z\, \mathcal A_- }{\sqrt{2}}\right)}  \right] - \sqrt{2}i\omega \left[ \frac{\sinh{\left( \frac{z\, \mathcal A_- }{\sqrt{2}}\right)}}{\mathcal A_-} -  \frac{\sinh{\left( \frac{z\, \mathcal A_+ }{\sqrt{2}}\right)}}{\mathcal A_+}  \right] \right|^2.
\end{equation}
The expression of the transition efficiency in \eqref{prob-2} is an exact one in the case of transition of photons into pseudoscalar particles in a transverse magnetic field and media and it is valid as far as \eqref{om-cond} is satisfied. This expression is valid both in the relativistic and non relativistic case. Expression \eqref{prob-2} generalizes the transition efficiency of photons into pseudoscalar previously obtained in the literature only in the relativistic case \cite{Raffelt88}. As we already discussed above, expression \eqref{prob-2} is a quantity bounded between zero and one as it should be from the way it is defined. So, if we fix some of the parameters, say $\omega, \bar B_x, z$ and $\Pi_{xx}$, then at least one of the parameters $g_{\phi\gamma}$ or $m_\phi$ cannot be completely arbitrary but must be constrained by the condition $P_{\gamma\phi}(z, \omega)\leq 1$.

As we will see in what follows it is quite useful to write $\mathcal S$ and $\mathcal A$ in \eqref{A-S-def} as follows
\begin{equation}\label{A-S-def-1}
\frac{\mathcal S}{2\omega^2} = \sqrt{\frac{g_{\phi\gamma}^2 \bar B_x^2}{\omega^2} +\frac{1}{4} \frac{\left(m_\phi^2 - \Pi_{xx}\right)^2}{\omega^4}}, \qquad \mathcal A_\pm = i\sqrt{2}\omega\sqrt{1 -\frac{1}{2}\left( \frac{m_\phi^2}{\omega^2} + \frac{\Pi_{xx}}{\omega^2} \pm \frac{\mathcal S}{\omega^2}\right)},
\end{equation}
where again we here consider only $\omega>0$ for simplicity. If we consider $\omega<0$ the expression for the transition efficiency remains invariant. By using expressions in \eqref{A-S-def-1} into \eqref{prob-2} we get the following general expression for the transition efficiency of photons in a magnetized plasma/gas for transverse external magnetic field to the direction of propagation of the wave
\begin{equation}\label{gen-prob}
\begin{gathered}
P_{\gamma\phi} (z, \omega) = \frac{g_{\phi\gamma}^2\omega^2 \bar B_x^2}{4\,\omega^2 g_{\phi\gamma}^2 \bar B_x^2 + (m_\phi^2 - \Pi_{xx})^2} \left\{ \left( \cos\left[\omega z\sqrt{1 -\frac{1}{2} \left( \frac{m_\phi^2}{\omega^2} + \frac{\Pi_{xx}}{\omega^2} + \frac{\mathcal S}{\omega^2}\right)}  \right]  - \cos\left[\omega z\sqrt{1 -\frac{1}{2} \left( \frac{m_\phi^2}{\omega^2} + \frac{\Pi_{xx}}{\omega^2} - \frac{\mathcal S}{\omega^2}\right)}  \right]   \right)^2  \right.\\ \left. + \left(  \left(\sqrt{1 -\frac{1}{2} \left( \frac{m_\phi^2}{\omega^2} + \frac{\Pi_{xx}}{\omega^2} + \frac{\mathcal S}{\omega^2}\right)} \right)^{-1}\sin\left[\omega z\sqrt{1 -\frac{1}{2} \left( \frac{m_\phi^2}{\omega^2} + \frac{\Pi_{xx}}{\omega^2} + \frac{\mathcal S}{\omega^2}\right)} \right] \right.\right.\\ \left.\left.  - \left(\sqrt{1 -\frac{1}{2} \left( \frac{m_\phi^2}{\omega^2} + \frac{\Pi_{xx}}{\omega^2} - \frac{\mathcal S}{\omega^2}\right)} \right)^{-1}\sin\left[\omega z\sqrt{1 -\frac{1}{2} \left( \frac{m_\phi^2}{\omega^2} + \frac{\Pi_{xx}}{\omega^2} - \frac{\mathcal S}{\omega^2}\right)} \right]  \right)^2 \right\}.
\end{gathered}
\end{equation}
Expression \eqref{gen-prob} is valid as far as the condition \eqref{om-cond-1} holds and it generalizes the transition efficiency found in the literature \cite{Raffelt88} to the case of non relativistic pseudoscalar particles in a dispersive magnetized medium. In \eqref{gen-prob} the pseudoscalar parameters cannot be arbitrary as we have already discussed above because of the fact that $P_{\gamma\phi}\leq 1$. Once that $\omega$ and $\bar B_x$ are fixed, we have to adjust the parameters $g_{\phi\gamma}$ and $m_\phi$ in order to have $P_{\gamma\phi}\leq 1$.

There are several useful hints that we can get from \eqref{gen-prob} and the conditions under which it has been derived. As we already discussed above we must have that $\mathcal A_\pm$ is purely imaginary for propagating fields in space that are satisfied for -$2\omega^2 + m_\phi^2 + \Pi_{xx} \pm \mathcal S<0$. The solution of the latter inequalities in terms of $g_{\phi\gamma}\bar B_x$ being the independent variable is given by
\begin{equation}\label{gen-cons}
0< g_{\phi\gamma}\bar B_x< \sqrt{\frac{(\omega^2-m_\phi^2)(\omega^2- \Pi_{xx})}{\omega^2}} \quad \text{for} \quad \omega^2> \{\Pi_{xx}, m_\phi^2 \}.
\end{equation}
So as we can see expression \eqref{gen-cons} gives a rather general constraint on $g_{\phi\gamma}\bar B_x$ from the requirement of propagating fields in space. Expression \eqref{gen-cons} tells us that we must have $\omega^2> \{\Pi_{xx}, m_\phi^2 \}$ as far as we have propagating fields and real and positive values of $g_{\phi\gamma}\bar B_x$.

It is quite useful to see as a matter of example of how expression \eqref{prob-2} or  \eqref{gen-prob} reduces in case when the mixing particles are relativistic, namely $\omega\gg m_\phi, |\Pi_{xx}|$ where $\Pi_{xx}$ essentially represents the effective mass of the photon state in a medium. In order to proceed further it is also very useful to calculate the order of magnitude of each term appearing in \eqref{A-S-def-1} where
\begin{equation}\label{defini-1}
\begin{gathered}
\left(\frac{g_{\phi\gamma} \bar B_x}{\omega}\right)^2= 8.78 \times 10^{-34} \left( \frac{g_{\phi\gamma}}{10^{-10} \text{GeV}^{-1}} \right)^2 \left( \frac{\bar B_x}{\text{T}} \right)^2 \left( \frac{10^{15} \text{Hz}}{\omega} \right)^2, \quad \left(\frac{m_\phi}{\omega} \right)^2 = 2.31 \times 10^{-12}  \left( \frac{m_\phi}{10^{-6}\text{eV}} \right)^2 \left( \frac{10^{15} \text{Hz}}{\omega} \right)^2,\\
\frac{|\Pi_{xx}|}{\omega^2} = 2.31 \times 10^{-20}  \left( \frac{|\Pi_{xx}|}{10^{-20}\text{eV}^2} \right) \left( \frac{10^{15} \text{Hz}}{\omega} \right)^2.
\end{gathered}
\end{equation}

By looking at expressions in \eqref{defini-1}, we can observe that each of them are much smaller than unity for reasonable values of the parameters $\omega, \bar B_x$ and $g_{\phi\gamma}$. Consequently we can keep only the first order terms in series expansion of $\mathcal A_\pm$ in \eqref{A-S-def-1}, while we keep exact the expression for $\mathcal S/(2\omega^2)$. By doing some lengthy calculations, we get the following expression for transition efficiency of photons into pseudoscalar particles for relativistic particles in the case when all quantities in \eqref{defini-1} are much less than unity
\begin{equation}\label{prob-3}
\begin{gathered}
P_{\gamma\phi}(z, \omega) \simeq   \frac{4\, g_{\phi\gamma}^2\omega^2 \bar B_x^2}{\mathcal S^2} \left[\sin^2\left( \frac{z \mathcal S}{4\,\omega}\right) - \frac{\mathcal S}{8\,\omega^2}\sin\left( \frac{z \mathcal S}{2\,\omega}\right) \sin\left[ 2 \omega z \left( 1- \frac{m_\phi^2 + \Pi_{xx}}{4\,\omega^2} \right)  \right] +\right. \\ \left.  \frac{m_\phi^2 + \Pi_{xx}}{2\,\omega^2} \sin^2\left( \frac{z \mathcal S}{4\,\omega}\right) \cos^2\left[ \omega z \left( 1- \frac{m_\phi^2 + \Pi_{xx}}{4\,\omega^2} \right)  \right] + \frac{\mathcal S^2}{16\,\omega^4}  \cos^2\left( \frac{z \mathcal S}{4\,\omega}\right) \sin^2\left[ \omega z \left( 1- \frac{m_\phi^2 + \Pi_{xx}}{4\,\omega^2} \right)  \right]  - \right. \\ \left. \frac{\mathcal S (m_\phi^2 + \Pi_{xx})}{32\,\omega^4}\sin\left( \frac{z \mathcal S}{2\,\omega}\right) \sin\left[ 2 \omega z \left( 1- \frac{m_\phi^2 + \Pi_{xx}}{4\,\omega^2} \right)  \right] + O\left( \frac{(m_\phi^2 + \Pi_{xx})^2}{16\, \omega^4} \right) \right],
\end{gathered}
\end{equation}
where the last $O$-term in \eqref{prob-3} is a smaller correction of the third term within parenthesis in \eqref{prob-3}. The expression for the transition efficiency in \eqref{prob-3} is a generalization of the transition efficiency of photons into pseudoscalar particles found in the literature by using approximate methods to solve the equations of motion, in the relativistic regime, for an incident linearly polarized plane electromagnetic wave or plane-like electromagnetic wave with constant or slowly varying amplitude with respect to the position $z$. Usually by using a WKB approximation for relativistic particles, in the literature only the first trigonometric term within the parenthesis in \eqref{prob-3} is present and the remaining terms are completely absent. The absence of the remaining terms is usually justified because their magnitudes are usually very small (much less than unity) and negligible for relativistic particles. However, these terms might be important in the case when the leading trigonometric term within the parenthesis in \eqref{prob-3} is identically to zero, namely when $\sin^2(z\mathcal S/(4\omega))=0$ that is satisfied for $z=4\,\omega\, n\,\pi/\mathcal S$ where $n$ is an integer number $n\in \bs Z$. In that case also the second, third, fifth and last terms within parenthesis in \eqref{prob-3} are identically to zero. The only surviving term is the fourth term within parenthesis in \eqref{prob-3} and the expression for the transition efficiency reduces to
\begin{equation}\label{prob-4}
\begin{gathered}
P_{\gamma\phi}(\omega; n) =   \frac{ g_{\phi\gamma}^2 \bar B_x^2}{4\, \omega^2} \sin^2\left[ \frac{2n\pi}{\mathcal S/(2\omega^2)} \left( 1- \frac{m_\phi^2 + \Pi_{xx}}{4\omega^2} \right)  \right]  \simeq \frac{ g_{\phi\gamma}^2 \bar B_x^2}{4\, \omega^2} \sin^2\left[ \frac{2n\pi}{\sqrt{\frac{g_{\phi\gamma}^2 \bar B_x^2}{\omega^2} +\frac{1}{4} \frac{\left(m_\phi^2 - \Pi_{xx}\right)^2}{\omega^2}}} \right] \quad \text{for}\,\quad z=\frac{4\pi\omega n}{\mathcal S}.
\end{gathered}
\end{equation}
Suppose for example that $m_\phi^2\gg |\Pi_{xx}|$ and also suppose that $g_{\phi\gamma}\bar B_x\ll m_\phi$ in \eqref{A-S-def-1}. In this case we can approximate $\mathcal S/(2\omega^2)\simeq (1/2) (m_\phi/\omega)^2$ in \eqref{A-S-def-1}. Under this approximation we have that expression \eqref{prob-4} reduces to
\begin{equation}\label{prob-5}
\begin{gathered}
P_{\gamma\phi}(\omega; n) = \frac{ g_{\phi\gamma}^2 \bar B_x^2}{4\, \omega^2} \sin^2\left[ \frac{4n\pi}{m_\phi^2/\omega^2} \right] \quad \text{for}\,\quad z=\frac{4\pi\omega n}{\mathcal S} \simeq \frac{1.63\times 10^6\,n \,(\text{m})}{\left( \frac{m_\phi}{10^{-6}\text{eV}} \right)^2 \left( \frac{10^{15} \text{Hz}}{\omega} \right)^2}.
\end{gathered}
\end{equation}

Let us consider for example that $n=1, m_\phi=10^{-6}$ eV and $\omega=10^{18}$ Hz (we should keep in mind that we are working under the condition $\omega\gg \omega_c$ for propagation in matter, see the approximations in Sec. \ref{sec:4}). For these values of the parameters, we would have $(m_\phi/\omega)^2= 2.31\times 10^{-18}\ll 1$ (we are within the limits of our approximations), $\sin^2\left[ \frac{4\pi}{m_\phi^2/\omega^2} \right] \simeq 0.55$ and the transition efficiency in \eqref{prob-5} would become
\begin{equation}\label{prob-6}
\begin{gathered}
P_{\gamma\phi}(\bar B_x, g_{\phi\gamma}) \simeq 1.2 \times 10^{-40} \left( \frac{g_{\phi\gamma}}{10^{-10} \text{GeV}^{-1}} \right)^2 \left( \frac{\bar B_x}{\text{T}} \right)^2 \leq 1.
\end{gathered}
\end{equation}
If we consider that $g_{\phi\gamma} \simeq 10^{-10}$ GeV$^{-1}$ and a strong magnetic field with strength $\bar B_x =10^{6}$ T, we would get $P_{\gamma\phi} \simeq 1.2 \times 10^{-28}$. This example tells us that if we would use the transition efficiency found by using approximation methods, we would get a transition efficiency that is exactly zero. However, if we use the full expression for the transition efficiency we would get a transition efficiency that is usually a very small quantity but not zero. If we add to this fact that the transition efficiency could be bigger or smaller depending on the values of the parameters and that for a practical purpose we multiply the value of the transition efficiency with very large numbers, then we can realize the importance of the exact expression for the transition efficiency. Of course, all said depends on particular situations and clearly, there are situations that the approximate expression for the transition efficiency is more than enough for practical purposes and the above corrections are irrelevant. 

Another interesting fact is that in the case of propagation in a magnetized vacuum and for arbitrary direction of the magnetic field, there is also the presence of the longitudinal component of the electric field in the case when the external magnetic field has a longitudinal component with respect to the direction of propagation. The appearance of this longitudinal state is in some sense analogous to the appearance of the longitudinal state of the electromagnetic radiation in a magnetized plasma. The electric field component of the electromagnetic radiation associated with this longitudinal state is given by $E_z(z, t)= -\partial_t A_z(z, t)=i\omega A_z(z, \omega) e^{-i\omega t}$ in the temporal gauge $A^0=0$ for a monochromatic wave. The magnitude of the electric field in the direction of propagation of the wave is given by using the expression for $A_z(z, \omega)$ in \eqref{sys-3} and if we consider only $\bar{\bs B}= (\bar B_x, 0, \bar B_z)$, we get
\begin{equation}\label{Lon-photon}
|E_z(z,t)| =\left| \frac{\mathcal G_- \mathcal G_+ \bar B_z}{2\sqrt{2} \omega^4\mathcal F\,\bar B_x} \right| \left| (\sqrt{2}\omega^2)^{-1} \left[ \cosh{\left( \frac{z\, \mathcal C_- }{\sqrt{2}\omega^2}\right)} -  \cosh{\left( \frac{z\, \mathcal C_+ }{\sqrt{2}\omega^2}\right)}  \right]   +   i\omega \left[ \frac{\sinh{\left( \frac{z\, \mathcal C_- }{\sqrt{2}\omega^2}\right)}}{\mathcal C_-} -  \frac{\sinh{\left( \frac{z\, \mathcal C_+ }{\sqrt{2}\omega^2}\right)}}{\mathcal C_+}  \right]\right| |E_x(0, t)|,
\end{equation}
where at $z=0$ we took $k=\omega$ for an incident plane monochromatic wave.

\section{Discussion and conclusions}
\label{sec:6}

In this work, I studied the mixing of the electromagnetic field with a pseudoscalar field in a magnetized vacuum and media. I studied the effects on the electromagnetic field in the case when the external magnetic field is transverse with respect to the direction of propagation in media and also for an arbitrary direction of the external magnetic field in vacuum.  To obtain the results found in this work, I solved exactly the equations of motion of the photon-pseudoscalar field in different situations in the presence of a constant magnetic field. One of the key aspects is that we had to reduce the second order partial differential equations to first order and then solve them for space independent coefficient matrix $M$. On doing this, I employed the same method used to solve the equations of motion of the graviton-photon mixing in an external magnetic field \cite{Ejlli:2020fpt}. The solution of the equation of motion of the photon-pseudoscalar particle mixing is solvable exactly for a constant magnetic field of arbitrary direction in a vacuum and a transverse magnetic field in magnetized media. For other configuration types, the solution of the equations of motion is usually extremely complicated because it is very difficult to analytically compute the matrix exponential of $(z-z_i)M$. 

In all cases studied in the work, in order to solve the equations of motion, I studied the case when the coefficient matrix $M$ is diagonalizable. The condition for $M$ to be diagonalizable, obviously comes with some constraints on the parameters that enter the theory as it has been discussed in Sec. \ref{sec:3} - \ref{sec:4}. So, the obtained  exact solution of the equations of motions are valid within the constraints imposed on the parameters. However, as we have seen the condition that $M$ is diagonalizable, physically implies that the electromagnetic and pseudoscalar fields must propagate in space, where the total energy must be bigger than the total effective masses of the particles either in media or in a vacuum.

One of the main results obtained in this work has been to find the exact solution of the equations of motion in some cases and compute the transformation efficiency or transition probability of photons into pseudoscalar particles. The expressions for the transition efficiencies, that have been found in different situations, generalize those previously found in the literature by using approximate WKB methods to solve the equations of motion. These expressions are those shown in \eqref{prob-1} in vacuum and that in \eqref{gen-prob} in matter. Both these expressions have never been found before in the literature and are valid in the non relativistic and relativistic cases. As a matter of example, we have seen in Sec. \ref{sec:5}, there are cases when the expression for the transition efficiency is exactly zero for some values of $z$ when approximate methods are used to compute it. However, when the equations of motion are solved exactly, there are additional trigonometric terms in the expression of the transition efficiency that are non zero. This fact might be important since there could be situations when these extra terms of the transition efficiency would be relevant if multiplied with large numbers such as, for example, the electromagnetic intensity of a given object etc. Also, the main expressions for the transition efficiencies (\eqref{prob-1} and \eqref{gen-prob}) have been found for arbitrary particle energies, as far as the propagation of the fields in space is concerned, contrary to those studies in the literature where the interacting particles have been usually assumed to be relativistic. The only conditions on the values of the total energy $\omega$ come from the diagonalizability of the matrix $M$, where usually $\omega$ must be bigger than the total effective masses of the particles in media for propagating fields as in expression \eqref{om-cond-1}.

Another important fact that has been neglected or ignored in the literature, apart from some studies \cite{Mikheev:1998bg} - \cite{Caputo:2020quz}, is that in the case when the external magnetic field has a longitudinal component with respect to the direction of propagation of the fields, it is present also a longitudinal component of the electric field (in a magnetized medium and in a magnetized vacuum), see Eq. \eqref{Lon-photon}. The appearance of this longitudinal electric field could be very important in laboratory searches of axions and/or axion-like particles. In principle, one can attempt to use the electromagnetic force associated with the longitudinal electric field on a charged object that can be put in the apparatus and then try to measure its displacement in response. Another important consequence of the appearance of the longitudinal electric field is that the rotation of the plane of the incident electromagnetic wave is not anymore necessary to occur in the plane perpendicular to the direction of propagation of the fields but it can be slightly tilted towards the direction of propagation. The appearance of the longitudinal electric field state and its applicability in laboratory searches of axions and/or axion-like particles depends on several factors and it could be very well that its appearance is irrelevant with respect to other techniques that try to find these weakly interacting particles.

The last thing that is worth mentioning and which goes in parallel with the conclusion of laboratory searches of graviton-photon mixing \cite{Ejlli:2020fpt}, is that our results have been found in the case when the propagation of the fields happens in open-like space and not in a confined cavity. If the photon-pseudoscalar particle mixing occurs in a confined cavity where the electromagnetic field could be reflected on the cavity walls, then it is necessary to look for the solution of the equations of motion for fields that depend on all spatial coordinates $x, y, z$ and not only $z$ as studied in this work.

\end{document}